\documentclass[aps,prb,twocolumn,floatfix,floats,showpacs,superscriptaddress,raggedbottom]{revtex4-1}
\usepackage{graphicx,latexsym}
\usepackage{dcolumn}
\usepackage{amsmath}
\usepackage{amssymb,bm}
\usepackage[normalem]{ulem}

\def\mb#1{\mbox{\boldmath$#1$}}

\def\eq#1{Eq.\ (\ref{#1})}
\def\fig#1{Fig.\ \ref{#1}}

\usepackage{hyperref}
\hypersetup{
    pdfnewwindow=true,       
    colorlinks=true,         
    linkcolor=blue,          
    citecolor=blue,          
    filecolor=magenta,       
    urlcolor=black           
}

\usepackage{booktabs}

\begin{document}

\title{Delocalization of electrons by cavity photons in\\ 
       transport through a quantum dot molecule}

\author{Nzar Rauf Abdullah}
\email{nra1@hi.is}
\affiliation{Science Institute, University of Iceland,
        Dunhaga 3, IS-107 Reykjavik, Iceland}

\author{Chi-Shung Tang}
\email{cstang@nuu.edu.tw}
 \affiliation{Department of Mechanical Engineering,
        National United University, 1, Lienda, Miaoli 36003, Taiwan}

\author{Andrei Manolescu}
 \affiliation{Reykjavik University, School of Science and Engineering,
              Menntavegur 1, IS-101 Reykjavik, Iceland}

\author{Vidar Gudmundsson}
\email{vidar@raunvis.hi.is}
 \affiliation{Science Institute, University of Iceland,
        Dunhaga 3, IS-107 Reykjavik, Iceland}

%

\begin{abstract}
We present new results on cavity-photon-assisted electron transport through 
two lateral quantum dots embedded in a finite quantum wire. 
The double quantum dot system is weakly connected to two leads and strongly coupled to a single quantized
photon cavity mode with initially two linearly polarized photons in the cavity.
Including the full electron-photon interaction, the transient current controlled by a plunger-gate in the central system
is studied by using quantum master equation.
Without a photon cavity, two resonant current peaks are observed in the range 
selected for the plunger gate voltage: The ground state peak, and the peak corresponding to the first-excited state.
The current in the ground state is higher than in the 
first-excited state due to their different symmetry.
In a photon cavity with the photon field polarized along or perpendicular to the transport direction, 
two extra side peaks are found, namely, photon-replica of the ground state and photon-replica of the first-excited state.
The side-peaks are caused by photon-assisted electron transport, with multiphoton absorption processes
for up to three photons during an electron tunneling process.
The inter-dot tunneling in the ground state can be controlled by 
the photon cavity in the case of the photon field polarized along the transport direction. 
The electron charge is delocalized from the dots by the photon cavity. 
Furthermore, the current in the photon-induced side-peaks can be strongly enhanced by increasing
the electron-photon coupling strength for the case of photons polarized along the transport direction.

\end{abstract}



\maketitle

%
%

\section{Introduction}

An opto-electronic device provides a different platform of electron transport, namely
photon-assisted transport (PAT)\cite{Kouwenhoven40.2001}. In the PAT, 
the energy levels of an electronic system have to match to photon frequency 
of a radiation source to control the electron motion. 
Therefore, the photon emission and the photon absorption processes 
play an essential role to enhance electron transport.\cite{Kouwenhoven.75.2003}
For that purpose, an electrostatic potential produced by a plunger-gate is applied to the electronic system 
to shift it's energy levels in and out of resonance.
The plunger-gate is widely used to control charge current \cite{Nzar.25.465302},
thermal current \cite{Tagani413:86_91}, photo-current \cite{Xu.87.035429}
and spin-dependent current \cite{Souza.84.115322} for various quantized systems coupled to photon radiation.

The PAT controlled by plunger-gate has been investigated
to study electrical \cite{Watzel99.192101} and optical \cite{Matutano84:041308,Kaczmarkiewicz114:183108}
properties of a double-quantum dot (DQD) system , in which the PAT can be used as a spectroscopic tool in two different regimes defined by
a zero \cite{Phys.Rev.Lett.76.1996}, and non-zero \cite{PhysRevB.53.1050} bias voltage.
At zero-bias voltage, the DQD works as a proper electron pumping device in which the photon absorption process leads to
electron tunneling producing a dc current.
In the non-zero bias voltage, both the photon absorption and the photon emission processes generate
a dc current. Recently, both regimes have been realized experimentally
in a DQD system at low temperature.\cite{Shang.103.2013,Shibata109.077401} 

The most important application of a DQD system in the quantum regime is intended for information storage 
in a quantum state,\cite{Imamog72.210} quantum-bits for quantum computing, \cite{Loss.57.1998,DiVincenzo.309.2005}
and quantum information processing in two-state system.\cite{Nielsen.2010} Recent experimental work has focused on using 
the two lowest energy states contributing to tunneling processes in a DQD  working as a two 
state system: The ground state resonance, and a photon-induced excited state resonance. They observed multiphoton absorption
processes up to the four-order contributing to the electron transport.\cite{Shibata109.077401}

Based on the above-mentioned considerations, we analyze PAT in serial double quantum dots embedded in a quantum wire.
The DQD system is connected to two leads and coupled to a photon cavity with linearly polarized photons 
in the $x$- and $y$-directions, where the transport along the quantum wire is in the $x$-direction. 
A quantum master equation (QME) formalism is utilized to investigate transient transport of electrons controlled
by the plunger-gate in the system without and with a single-photon mode.~\cite{Nzar.25.465302}
Generally, there are two types of QME when characterized according to memory effects, energy-dependent coupling, 
and the system-leads coupling strength: The Markovian and the non-Markovian QME.
In the case of the Markovian approximation, the system-leads coupling is assumed weak and independent of energy,
memory effect are ignored and most commonly a steady state is sought.~\cite{Vaz81.085315, Gurvitz53.15932,Harbola74.235309,Kampen2.2001}
In the non-Markovian approach, the system is energetically coupled to the leads including memory effect
in the system.~\cite{Bednorz101.206803,Braggio96.026805,Emary76.161404}
Since we are interested in studying transient transport of electrons in a regime with possible resonances,
the non-Markovian model is used in our system.~\cite{Vidar11.113007}

In addition, we assume the DQD system to be connected to the leads through a non-zero or small bias window, 
where the two lowest energy states of the QDQ system can be isolated in the bias window: The ground state and the first-excited state.
Our model of the DQD system can be seen as a qubit. 
In which the states $\lvert 0\rangle$ and $\lvert 1\rangle$ can be represented in terms of the ground state and
the first-excited state.
We will show how the single-photon mode affects the electron transport through both states when located in the bias window
and demonstrate the role of photon activated states in the transient current. 
The double serial quantum dot is essential here: The two lowest single-electron states of the dot
molecule have very different symmetry. The ground state has a symmetric wavefunction, but the excited
state has an antisymmetric one. The conduction through the ground state is thus higher than through 
the excited one. The ``inter-dot tunneling'' can be influenced by a photon mode polarized in the 
transport direction, thus strongly modifying the conduction through the photon replicas of the states
in a photon cavity. The nontrivial details of this picture will be analyzed in this paper reminding
us that the effects rely on the geometry of the system and states beyond the ground state and the 
first excited one. 

The rest of the paper is organized as follows. In Sec.~\ref{Sec:II}
we introduce the model to describe the electron transport through 
a DQD embedded in a quantum wire connected to two leads and a photon cavity.
Section. \ref{Sec:III} contains two subsections, the system without and with the photon cavity.
In the absence of the photon cavity, the transient current through the system controlled by the plunger-gate is demonstrated
in the presence of the electron-electron interactions in the DQD system. In the photon cavity, 
the photon-assisted electron transport in 
the system is presented for a system initially with no electron, but with two linearly polarized photons in the single-photon mode.
Finally, conclusions are provided in Sec.~\ref{Sec:IV}.

\section{Model and Computational methods}\label{Sec:II}
\begin{figure}[htbq!]
 \includegraphics[width=0.48\textwidth,angle=0]{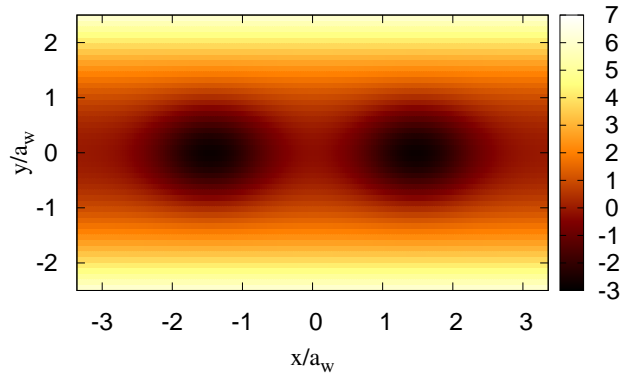}
 \caption{(Color online)
      Schematic diagram depicts the potential representing the DQD
      embedded in a quantum wire with parameters $B = 0.1~{\rm T}$,
      $a_{w} = 23.8~{\rm nm}$, and $\hbar \Omega_0 = 2.0~{\rm meV}$.}
      \label{DQD}
\end{figure}

The aim of this study is to model a photon-assisted electron transport
in a DQD system connected to two identical electron reservoirs (lead) and
coupled to a single photon mode in a cavity.
Our first step is to look at the central system, in which 
electrons are confined in two dimensions. We assume
a finite quantum wire with hard-wall ends at $x$ = $\pm L_{x}/2$ 
with length $L_x=165$~nm. It is parabolically confined in the $y$-direction
(perpendicular to the transport direction) 
with transverse confinement energy  $\hbar\Omega_0 = 2.0$~meV.
The embedded quantum dots are modeled by two identical Gaussian potentials in the quantum wire defined as 
\begin{equation}\label{potential}
      V_{\rm DQD}(x,y) =  \sum_{i = 1}^2 V_i\; \exp{\left[ -\beta_{i}^2\left( (x-x_i)^2 + y^2
      \right)    \right]},
\end{equation}
with quantum-dot strength $V_{1,2} = -2.8$ meV, $x_{1} = 35$~nm, $x_{2} = -35$~nm, 
and $\beta_{1,2}$ = $5.0\times10^{-2}~{\rm nm^{-1}}$ such
that the radius of each quantum-dot is $R_{\rm QD} \approx 20$~nm.
A sketch of the DQD system under investigation is
shown in \fig{DQD}. We should mention that the distance between the dots is $L_{\rm DQD} = 35~{\rm nm} \simeq 1.47a_w$,
and each dot is $25~{\rm nm} = 1.05a_w$ away from the nearest lead, where
$a_w$ is the effective magnetic length.

The DQD system is in a rectangular photon cavity with a single photon mode.
The photons in the single photon mode are linearly polarized in the
$x$- or $y$-directions, meaning that the photon polarization in the cavity is assumed to be parallel  
or perpendicular to the transport direction with respect to the electric field
\begin{equation}
      \mathbf{A_{\rm ph}} = A_{\rm ph}
       \left( a+a^{\dagger} \right)\mathbf{\hat{e}},
\end{equation}
where $A_{\rm ph}$ is the amplitude of the photon vector potential, $a^{\dagger}(a)$ are
the creation (annihilation) operators for a photon, respectively, and $\mathbf{\hat{e}}$ determines 
the polarization with
\[ \mathbf{\hat{e}} = \left\{ 
  \begin{array}{l l}
    (e_x,0), & \quad \text{ TE$_{011}$}\\
    (0,e_y), & \quad \text{ TE$_{101}$},
  \end{array} \right.\]
where TE$_{011}$ (TE$_{101}$) indicates the parallel (perpendicular) 
polarized photon in the transport direction, respectively.

In the following sections, we shall couple the DQD system to both the photon cavity and
the leads.

\subsection{DQD system coupled to Cavity}

We consider the closed DQD system to be strongly coupled to a photon cavity.
The many-body (MB) Hamiltonian  
\begin{equation}
H_\mathrm{S} = H_\mathrm{DQD} +  H_\mathrm{Cavity} + H_\mathrm{Int}
\label{HS-e-ph}
\end{equation}
consists of the Hamiltonian for the closed DQD system with the 
electron-electron interaction $H_\mathrm{DQD}$,
the free photon cavity Hamiltonian $H_\mathrm{Cavity}$, and the Hamiltonian for the  
electron-photon interaction $H_\mathrm{Int}$.

The DQD system (and the external leads) is placed in an external uniform perpendicular magnetic field $B\hat{z}$
in the $z$-direction defining an effective lateral confinement length $a_w = (\hbar /m^* \sqrt{(\omega^{2}_c + \Omega^2_0)})^{1/2}$,
where the effective electron mass is $m^*=0.067m_e$ for GaAs material and
$\omega_c = e B/m^*c$ is the cyclotron frequency. 
The Hamiltonian for the DQD system in a magnetic field including the 
electron-electron interaction can be written as
\begin{eqnarray}
           H_\textrm{DQD} &=& \sum_{i,j} \langle\psi_i \lvert\left\{  \frac{\bm{\pi}_e^2}{2m^*}
              + V_\mathrm{DQD} + eV_\mathrm{pg} \right\} \lvert \psi_j\rangle \delta_{i,j} d_i^{\dagger} d_j  \nonumber \\
           && + H_\mathrm{Coul} +   H_\mathrm{Z},
           \label{H_S}
\end{eqnarray}
where $|\psi_i\rangle$ stands for a single-electron states, $V_{\mathrm{pg}}$ is the the plunger gate
potential that shifts the energy levels of the DQD system with respect to the chemical potentials of the leads,
$d^\dagger_{i}$ ($d_{j}$) is an operator that creates (annihilates) an electron in the DQD system, respectively.
Moreover, the canonical momentum is $\bm{\pi}_e= \bm{p}+\frac{e}{c}\mathbf{A}_{\mathrm{ext}}$ 
with the kinetic momentum operator $\bm{p}$, and the vector potential in the Landau gauge 
$\mathbf{A}_{\mathrm{ext}}$ =  ($0,-By,0$). The electron-electron interaction in the central system 
is given by
\begin{equation}
               H_\mathrm{Coul} = \frac{1}{2}\sum_{ijrs} V_{\textrm{ijrs}}
               d_i^{\dagger}d_j^{\dagger}d_sd_r,
\end{equation}
with the Coulomb matrix elements $V_{\textrm{ijrs}}$.\cite{Abdullah52.195325}
The characteristic Coulomb energy is $E_{\rm C} = e^2/(2\varepsilon_r a_w) \approx 2.44~{\rm meV}$ 
at $B=0.1$~T with $a_{w} = 23.8~{\rm nm}$ and $\varepsilon_r=12.4$, the dielectric constant of GaAs. 
The characteristic Coulomb energy is greater than the thermal energy of the leads.
An exact numerical diagonalization method is used here for solving the Coulomb interacting many-electron 
(ME) Hamiltonian in a truncated Fock space
to obtain the ME energy spectrum of the DQD system.\cite{Yannouleas70.2067}
The Zeeman Hamiltonian shown in the third part of \eq{H_S} 
describes the interaction between the external magnetic field 
and the magnetic moment of an electron
\begin{equation}
H_{\rm Z}= \pm \frac{g^\ast \mu_B}{2}  B,
\end{equation}
where $\pm$ stands for $z$-spin components, 
$\mu_B = e\hbar/2m_{\rm e}c$ is the Bohr magneton, and 
the effective Lande $g$-factor is $g^\ast = -0.44$ for GaAs.

In order to investigate photon-assisted electron transport in the DQD system,
the electronic system is coupled to a photon cavity. The Hamiltonian of the free photon cavity 
is given by
\begin{equation}
 H_\textrm{Cavity} = \hbar\omega_{\textrm{ph}} \hat{N}_{\rm ph},
\end{equation}
where $\hbar \omega_{\textrm{ph}}$ is the energy of the single mode in the cavity, and $\hat{N}_{\rm ph} =
a^{\dagger} a$ is the photon number operator.
The interaction of the single quantized electromagnetic mode with the electronic system is 
described by the Hamiltonian including both the diamagnetic and the 
paramagnetic interactions of photons and electrons
\begin{eqnarray}
      H_\textrm{Int} &=& g_{\rm ph}\sum_{ij}d_i^{\dagger}d_j\; g_{ij}
      \left\{a + a^\dagger\right\} \nonumber \\
      &&+\frac{g_{\rm ph}^2}{\hbar\Omega_w} \sum_{i}d_i^{\dagger}d_i
      \left[  \hat{N}_{\rm ph} + \frac{1}{2}\left( a^\dagger a^\dagger + aa  + 1 \right)\right]
\label{H_Int}
\end{eqnarray}
herein, $g_{\rm ph} = e A_{\rm ph} \Omega_wa_w/c$ is the electron-photon coupling strength,
and $g_{ij}$ are dimensionless electron-photon coupling matrix elements.\cite{Vidar85.075306} 

Finally, the MB system Hamiltonian $H_{\rm S}$ is diagonalized in a
MB Fock-space $\{ |\breve{\alpha}\rangle\}$ to obtain the MB energy spectrum
of the DQD system coupled to the photon cavity.\cite{Vidar61.305} 
The diagonalization builds a new interacting MB state basis $\{|\breve{\nu})\}$, 
in which $|\breve{\nu}) = \sum_{\alpha}{\cal W}_{\mu\alpha}|\breve{\alpha}\rangle$ 
with ${\cal W}_{\mu\alpha}$ being a unitary transformation matrix.
The unitary transformation is used to convert the QME and the physical observables 
from non-interacting MB basis to the interacting MB basis. 

\subsection{DQD system connected to leads}

The DQD system is connected to two semi-infinite leads
with the same width. The chemical potential of the lead $l$ is $\mu_l$, 
with $l\in\{L,R\}$ being the left $L$ and the right $R$ lead. The
the Fermi function in the isolated lead $l$ before coupling to the central system is
$f_l\left( \epsilon_l(\mb{q}) \right) = \{\exp[\epsilon_l(\mb{q})-\mu_l]+1\}^{-1}$, where 
$\epsilon_l$ is the SE subband energy of the lead $l$ ($\mb{q}$ is the momentum dummy index.~\cite{Nzar.25.465302}) 
found from the non-interacting ME Hamiltonian of lead $l$ 
\begin{equation}
   H_l = \int d{\mb{q}}\, \epsilon_l(\mb{q}) {c^\dagger_{{\mb{q}}l}}
   c_{{\mb{q}}l},
\end{equation}
with ${c^\dagger_{{\mb{q}}l}}$ ($c_{{\mb{q}}l}$)
the electron creation(annihilation) operator in
lead $l$, respectively.\cite{Jinshuang128.1234703} 

In order to instigate electron transport between the subsystems,
the DQD system is coupled to the leads with energy dependent coupling coefficients
reflecting the geometry of the system
\begin{equation}
 T_{{\mb{q}i}l} =
 \int d\mathbf{r} d\mathbf{r^{\prime}} \psi_{\mb{q}l}(\mathbf{r}')^*
 g_{\mb{q}il} (\mathbf{r},{\bf r'}) \psi^\mathrm{S}_i({\bf r}).
 \label{Tlqn}
\end{equation}
An electron may be transferred from a state $|\mb{q}\rangle$ 
with the wavefunction $\psi_{\mb{q}l}(\mathbf{r}')$ in the leads 
to a SE state $|i\rangle$ with the SE wavefunction $\psi^\mathrm{S}_i({\bf r})$ in the DQD system
and vice versa, where the coupling function is $g_{\mb{q}il} ({\bf r},{\bf r'})$.\cite{Vidar11.113007}
The coupling coefficients are utilized to construct a time-dependent coupling Hamiltonian 
in the second quantization language 
\begin{equation}
      H_{{\rm T}l}(t)= \chi_l(t) \sum_{i}\int d{\mb{q}}\, \left[  {c^\dagger_{{\mb{q}}l}} T_{{\mb{q}i}l} d_i
      +   d^\dagger_i (T_{{i\mb{q}l}})^* c_{{\mb{q}}l}\right], 
      \label{H_T}
\end{equation}
with a time-dependent switching function $ \chi_l(t) = 1 - 2\{\exp [\alpha_l (t-t_0)] + 1\}^{-1}$ 
with $\alpha_l = 0.3$~${\rm ps}^{-1}$ being a switching parameter.

After the DQD system is coupled to the leads at $t= 0$, we calculate
the time evolution of the electrons and photons using the  
density operator and its equation of motion, the Liouville-von Neumann (Lv-N) equation 
$i\hbar \dot{W}(t)= \left[H(t),W(t)\right]$ for the whole
system. As this can not be accomplished we resort to using a 
projection formalism taking a trace over the Hilbert space of the leads
introducing the reduced density operator  
\begin{equation}
\rho(t)={\rm Tr}_\mathrm{L}{\rm Tr}_\mathrm{R} W(t)
\end{equation}
with $\rho(t_0) = \rho_\mathrm{S}$\cite{Haake3.1723}
and the condition that $W(t<t_0) = \rho_\mathrm{L}\rho_\mathrm{R}\rho_\mathrm{S}$ 
is the density operator of the total system before the coupling
with $\rho_\mathrm{S}$ being the density operator of the isolated DQD system.~\cite{Breuer2002}
The density operator of the leads before the coupling 
is $\rho_l= \exp[{-\beta (H_l-\mu_l N_l)}]/{\rm Tr}_l \{\exp[{-\beta(H_l-\mu_l N_l)}]\}$,
where $\beta = 1/k_{B}T_l$ is the inverse thermal energy, and
$N_l$ is the number operator for electrons in the lead $l$,\cite{Jinshuang128.1234703}

The time-dependent mean charge in the central system, the current in the leads are 
calculated from the reduced density operator as has been detailed in earlier 
publications.\cite{Nzar.25.465302,Vidar61.305}

\section{Numerical Results}\label{Sec:III}

In this section, we discuss the transport properties through the DQD system controlled by
plunger-gate voltage in both cases without a photon cavity and with 
$x$- or $y$-polarized photons in a cavity.

In order to obtain the PAT, the system has to satisfy the following
conditions: the MB energy level spacing has to be greater than the thermal energy $\Delta E_{\rm MB} > k_B T$,
and the MB energy level spacing has to be smaller or equal to the photon energy 
$\Delta E_{\rm MB} \leq \hbar \omega_{\rm ph}$.\cite{Kouwenhoven73.3443}
Initially the temperature of the central system is assumed to be $T=0$~K, 
and the leads are at $T=0.01$~K initially. Other physical parameters 
of the system are presented in table \ref{table1}.
\begin{table}[h!]
\begin{tabular}{l*{2}{c}r}
Quantity                                      & Typical parameter  \\
\hline
Thermal energy in leads ($k_B T$)             & $\approx$ 8.617$\times$10$^{-4}$ meV    \\
Characteristic Coulomb energy ($E_{\rm C}$)   & $\approx 2.44$~meV    \\
Photon energy  ($\hbar\omega_{\textrm{ph}}$)    & $= 0.25$~meV   \\
\end{tabular}
\caption{Characteristic energy scales of the system}\label{table1}
\end{table}

In addition, we assume the external magnetic field to be $B = 0.1$~T with the effective
lateral confinement length $a_w = 23.8$~nm and initially no electron is in the DQD system.

\subsection{The DQD system without the photon cavity}

In this section, the properties of the electron transport through the DQD system are presented
in the absence of the photon cavity in order to establish a comparison for later results for
transport through the system inside a cavity.  

\begin{figure}[tbhq]
 \includegraphics[width=0.23\textwidth,height=0.24\textheight,angle=0,viewport=0 5 200 240,clip]{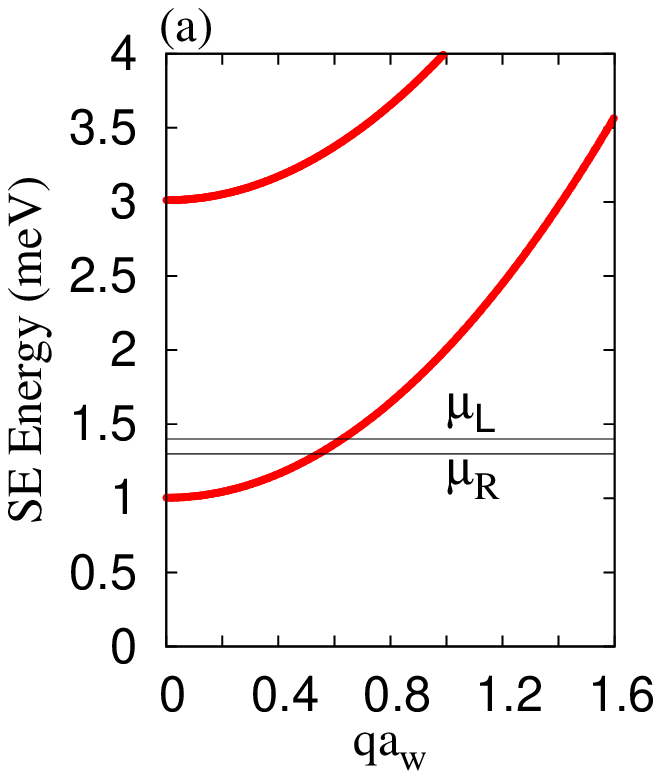}
 \includegraphics[width=0.23\textwidth,height=0.24\textheight,angle=0,viewport=0 3 200 240,clip]{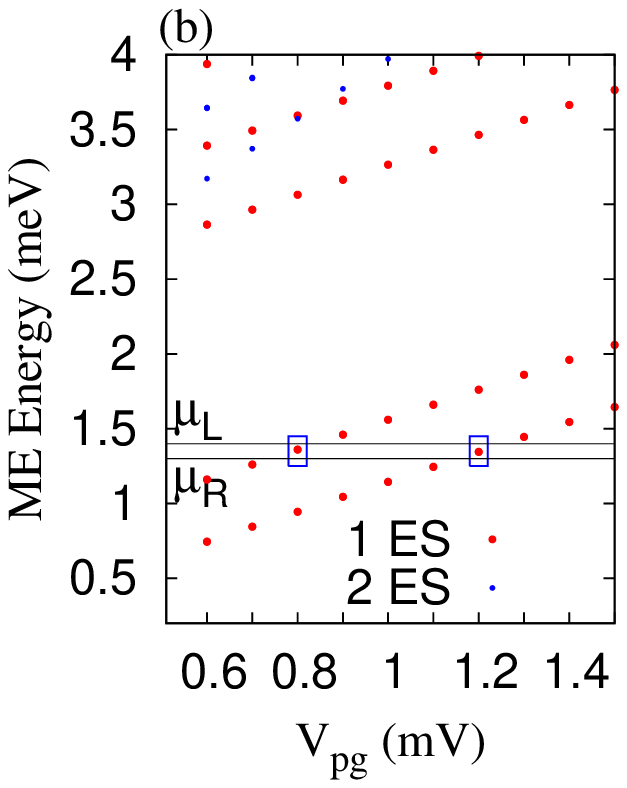}
 \caption{(Color online)
Energy spectra in the case of no photon cavity with magnetic field $B=0.1$~T. (a) SE
energy spectrum in the leads (red) is plotted as a function of scaled wave number $q a_w$, where
the chemical potentials are $\mu_L = 1.4\ {\rm meV}$ and $\mu_R = 1.3\ {\rm meV}$
(black). (b) ME energy spectrum in the central system as a function of plunger gate
voltage $V_{\rm pg}$ including SE states  (1ES, red dots) and two electron states (2ES,
blue dots). The SE state in the bias window is almost doubly degenerate due to the
small Zeeman energy.} \label{ME_e_e}
\end{figure}

Figure \ref{ME_e_e}(a) shows the energy spectrum of the leads versus
wave number $qa_w$. The horizontal black lines are chemical potentials
of the left lead $\mu_L$ and the right lead $\mu_R$. The chemical potentials are considered to be 
$\mu_L = 1.4\ {\rm meV}$ and $\mu_R = 1.3\ {\rm meV}$, implying a small bias voltage $\Delta\mu = 0.1$~meV.
Therefore, the first subband in the parabolic energy spectrum becomes the most active subband 
contributing to the electron transport process in the energy range [$1.3,1.4$]~meV.

In \fig{ME_e_e}(b), the ME energy spectrum of the DQD-system as a function of applied
plunger-gate voltage $V_{\rm pg}$ is shown. The energies of two-electron states $N_e = 2$ (2ES, blue dots)
are higher than the SE states $N_e = 1$ (1ES, red dots) due to the electron-electron interaction.
In the absence of the photon cavity, two resonant SE states are situated in the bias window
for the range of plunger gate voltage selected here, namely, the  
ground state resonance and first-excited state resonance (blue squared dots).

The almost degenerate two spin states of the single-electron ground state are $|2)$ and $|3)$ with energies 
$E_{\rm 2} = 1.343$~meV and $E_{\rm 3} = 1.346$~meV.
These two states get into resonance with the first-subband of the leads at $V_{\rm pg}^{\rm G} = 1.2$~mV,
where the superscript $\rm G$ refers to the ground state.
By tuning the plunger-gate voltage, the two spin states of the first-excited state
$|4)$ and $|5)$ with energies $E_{\rm 4} = 1.358$~meV 
and $E_{\rm 5} = 1.361$~meV contribute to the electron transport at $V_{\rm pg}^{\rm FE} = 0.8$~mV,
where the superscript $\rm FE$ stands for the first-excited state.

\begin{figure}[tbqh]
 \includegraphics[width=0.5\textwidth,angle=0]{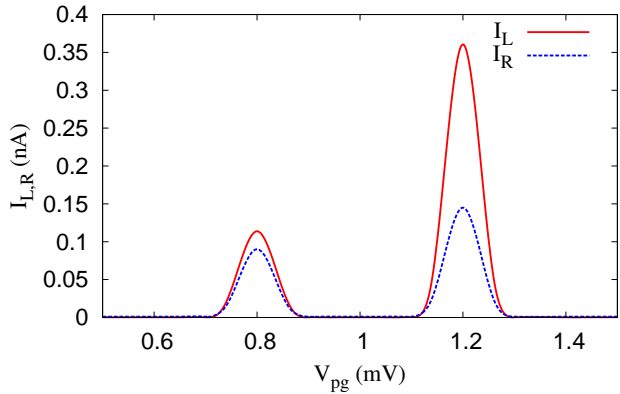}
\caption{(Color online) The left current $I_L$ (red solid) and right current $I_R$ (blue dashed) are plotted as a function of
plunger gate voltage $V_{\rm pg}$ at time $t = 220$~ps in the case of no photon cavity.
Other parameters are $B = 0.1~{\rm T}$ and $\Delta \mu =0.1~{\rm meV}$.
 } \label{I_Vpg_e_e}
\end{figure}

Figure \ref{I_Vpg_e_e} displays the left current $I_L$ (red solid) and the right current $I_R$ (dashed blue)
through the DQD system.
We notice two resonance peaks in the currents: The ground state peak at $V_{\rm pg}^{\rm G} = 1.2$~mV and
the first-excited state peak at $V_{\rm pg}^{\rm FE} = 0.8$~mV. The reason for the two current peaks is resonance of 
the SE states in the DQD system with the first subband energy of the leads. An electron in the first-subband of the left lead
may tunnel to the state $|2)$ or $|3)$ of the DQD system and subsequently tunnel out to the right lead.
Consequently the ground state peak is observed at $V_{\rm pg}^{\rm G}=1.2$~mV. In addition, the first-excited state peak reflects a
resonance with the states $|4)$ and $|5)$ at plunger-gate potential $V_{\rm pg}^{\rm FE}=0.8$~mV.

\begin{figure}[htbq]
 \includegraphics[width=0.23\textwidth]{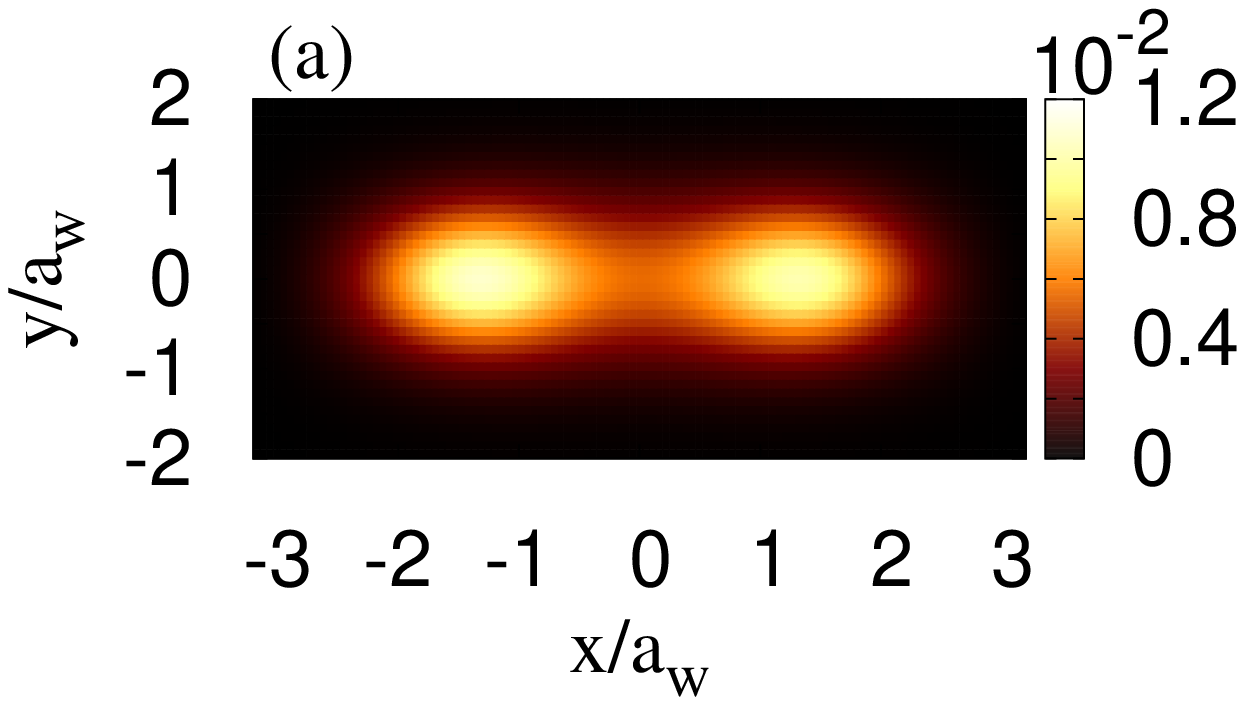}
 \includegraphics[width=0.23\textwidth]{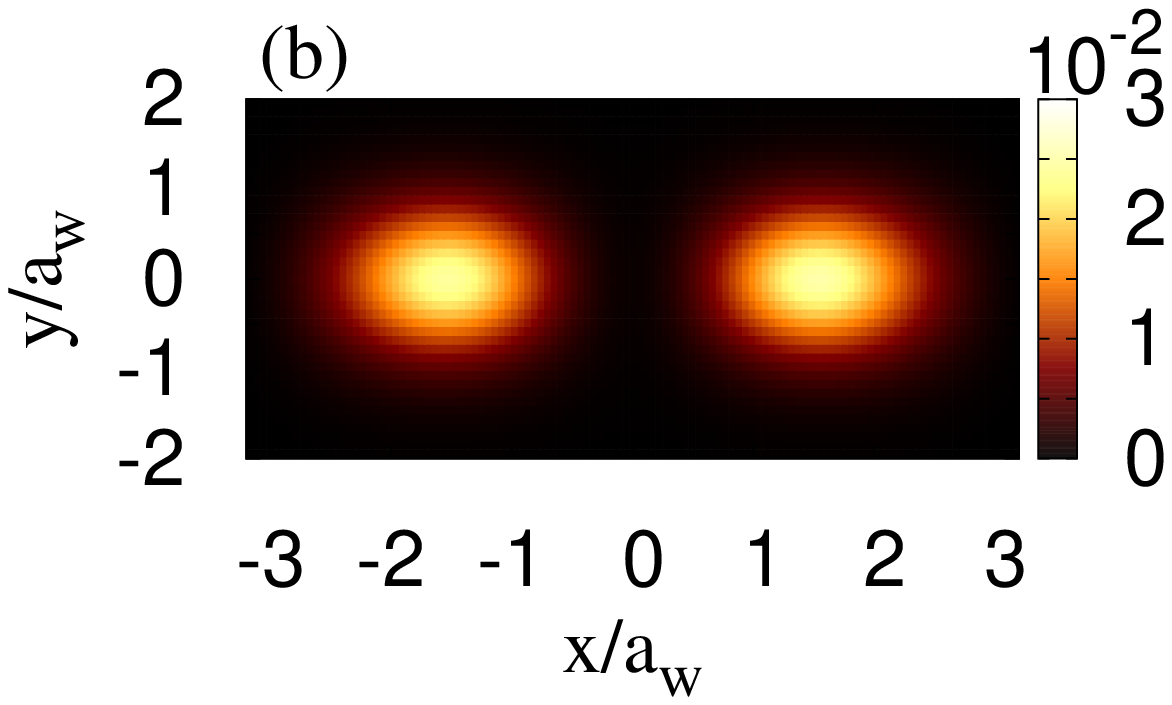}
 \caption{(Color online) The charge density distribution at 
$t = 220$~ps in the ground state peak (a) and first-excited state peak (b) shown
in \fig{I_Vpg_e_e} in the case with no photon cavity. Other parameters are
$B=0.1$~T, $a_{w}$ = $23.8~{\rm nm}$, $L_x$ = $165$~nm = $6.93 a_w$, and $\Delta \mu
=0.1~{\rm meV}$.} \label{Q_e-e}
\end{figure}

Figure \ref{Q_e-e} shows the charge density distribution in the DQD system at time $t = 220$~ps
(after the initial transient, close to a steady state) in the ground state peak (a), and the first-excited state peak (b).
In the case of the ground state peak at $V_{\rm pg}^{\rm G} = 1.2$~mV, the electron state accumulates in the dots
with a strong inter-dot tunneling. Therefore, the left and right currents increase in the system.
But in the case of first-excited state peak at $V_{\rm pg}^{\rm FE} = 0.8$~mV, 
the electron state is strongly localized in the dots 
without much tunneling between the dots. Thus the tunneling
between the dots is sufficiently suppressed and the current drops as shown in \fig{I_Vpg_e_e}.

\subsection{$x$-photon polarization (TE$_{011}$ mode)}

In this section we analyze the electron transport through the DQD system in the presence of an $x$-polarized 
single-photon mode with initially two photons in the cavity.
The photons in the cavity can excite electrons in the DQD system and enhance the electric current, similar
to the ``classical'' PAT case.\cite{PhysRevB.53.1050}
The condition for PAT involving $N_{\rm ph}$ photon(s) is 
$|E_i - E_f| = N_{\rm ph}\hbar\omega_{\rm ph}$,\cite{Platero.395.2004} where $E_i$($E_f$) is
the highest possible initial (lowest possible final) MB energy level of the DQD system, respectively.\cite{Kouwenhoven73.3443}
We vary the applied plunger-gate to match $|E_i - E_f|$ to the photon energy, thus
the PAT is an active process in the system.

\begin{figure}[tbhq]
 \includegraphics[width=0.3\textwidth]{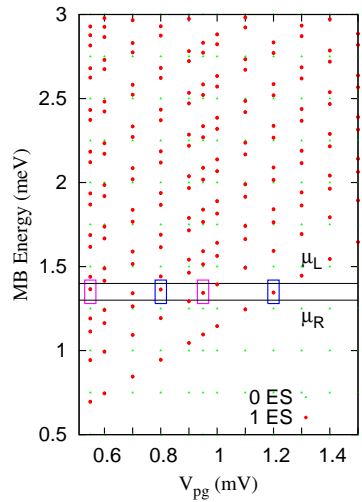}
 \caption{(Color online) MB Energy spectrum versus the plunger gate voltage $V_{\rm pg}$
 in the case of $x$-polarized photon field, where 0ES indicates zero electron states (Ne = 0, green dots),
and 1ES stands for single electron states (Ne = 1, red dots). 
Other parameters are $B=0.1$~T, $\Delta\mu = 0.1$~meV, and $\hbar
\omega_{\rm ph} = 0.25~{\rm meV}$. } \label{MBE_Xp}
\end{figure}

Figure \ref{MBE_Xp} shows the MB energy spectrum of the DQD system including
the photons with zero-electron states $N_e = 0$ (0ES, green dots)
and SE states $N_e = 1$ (1ES, red dots).
In addition to the former states at $V_{\rm pg}^{\rm G} = 1.2$~mV and  $V_{\rm pg}^{\rm FE} = 0.8$~mV, 
two extra active MB-states are observed in the presence of the photon cavity at
$eV_{\rm pg}^{\rm G_{\gamma};FE_{\gamma}} = eV_{\rm pg}^{\rm G;FE} - \hbar \omega_{\rm ph}$ in the bias window
(pink squared dots), where the photon energy is $\hbar\omega_{\rm ph} = 0.25$~meV, and
$\rm G_{\gamma}(\rm FE_{\gamma})$ stands for the photon-replica of the ground state(first-excited state), respectively.
We notice that all states in the bias window are SE states containing only one-electron $N_{\rm e} = 1$.

\begin{figure}[tbhq]
 \includegraphics[width=0.5\textwidth]{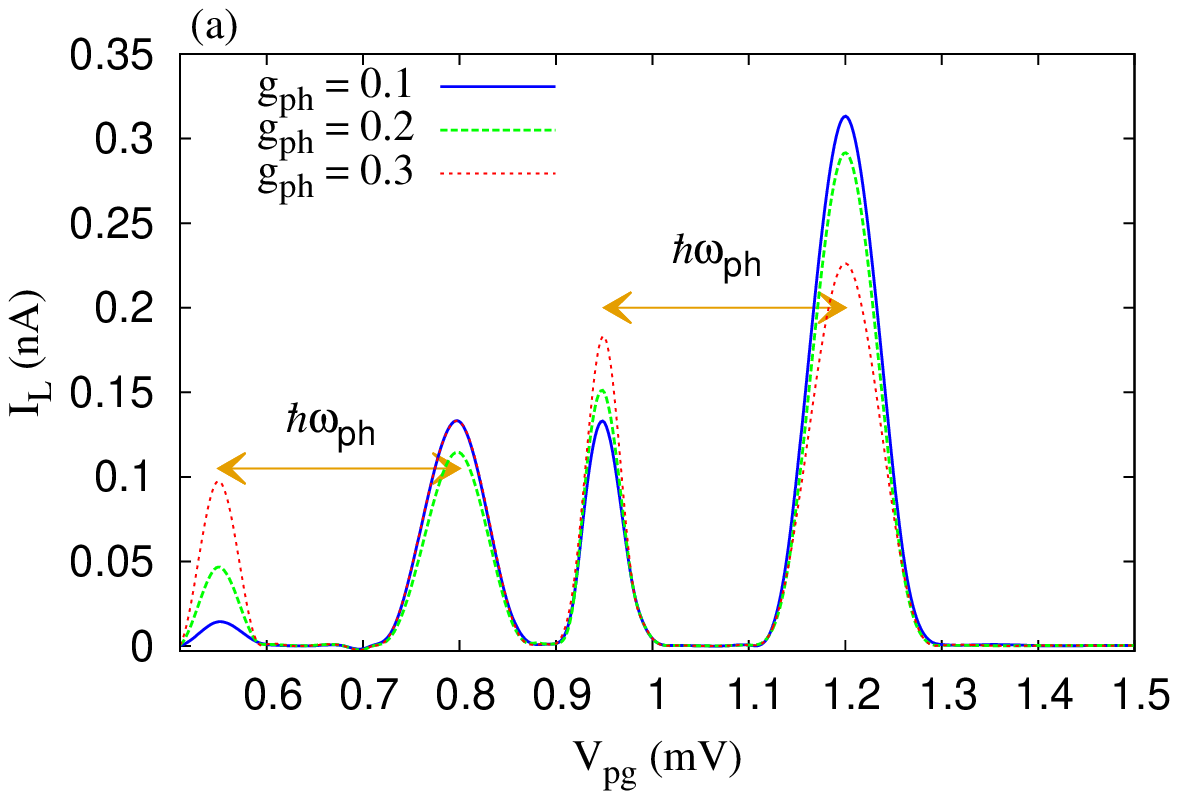}\\
 \includegraphics[width=0.5\textwidth]{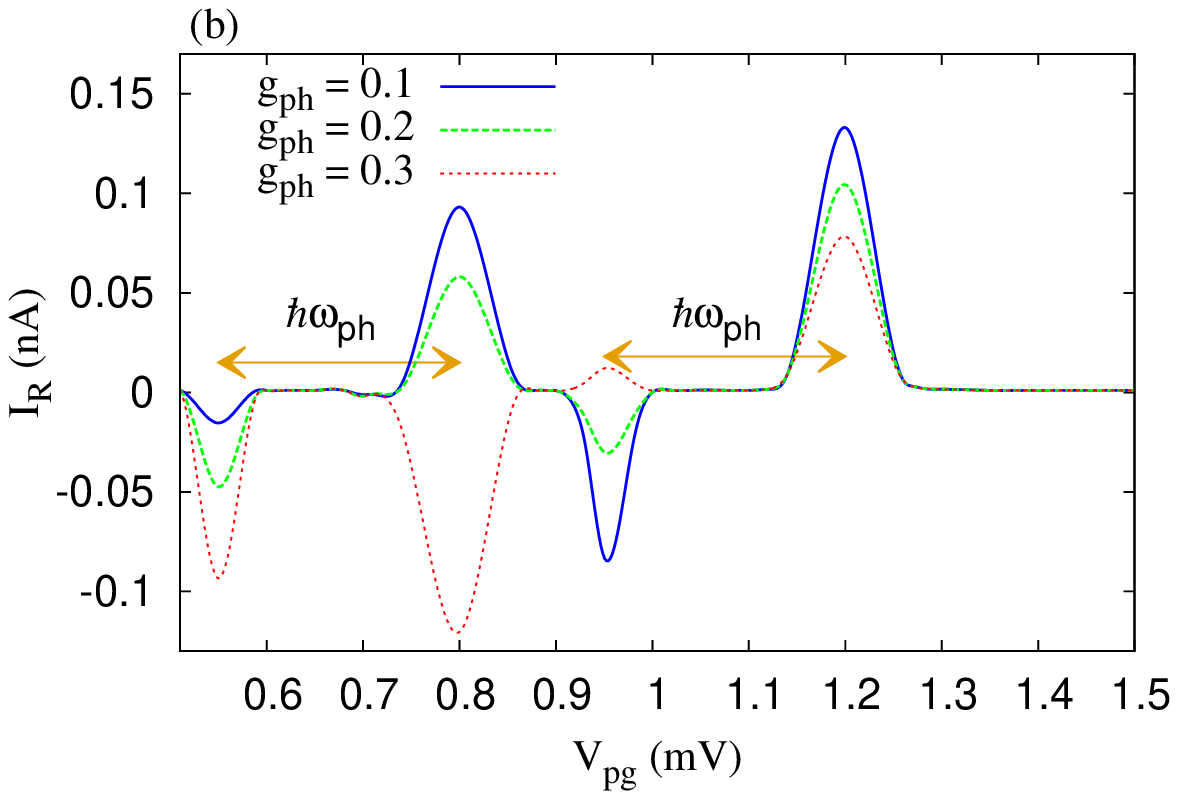}
\caption{(Color online) The left current $I_L$ (a), and the right current $I_R$ (b)
 versus the plunger gate voltage $V_{\rm pg}$ in the case of $x$-polarized photon field at time $t = 220~{\rm ps}$
 with different electron-photon coupling strength:
 $g_{\rm ph} = 0.1$~meV (blue solid), $0.2$~meV (green dashed),
 and $0.3$~meV (red dotted).
 Other parameters are $\hbar\omega_{ph} = 0.25~{\rm meV}$, $\Delta \mu =0.1~{\rm meV}$, and $B = 0.1~{\rm T}$.
 }
 \label{I-Xp}
\end{figure}

Figure \ref{I-Xp} displays the left current $I_L$ (a) and the right current $I_R$ (b)
as a function of the plunger-gate voltage $V_{\rm pg}$ in the presence of the $x$-polarized photon field at
time $t = 220$~ps for different electron-photon coupling strength  $g_{\rm ph} = 0.1~{\rm meV}$ (blue solid),
$0.2~{\rm meV}$ (green dashed), and $0.3~{\rm meV}$ (red dotted).
The positive value of the left current indicates electrons tunneling from the left lead to the DQD system, 
while the negative value of the right current denotes electrons tunneling from the right lead to the DQD system
and vise versa.

In the absence of the photon cavity, two main-peaks are found at $V_{\rm pg}^{\rm FEM} = 0.8$~mV 
and $V_{\rm pg}^{\rm GM} = 1.2$~mV as shown in \fig{I_Vpg_e_e}.
In the presence of the photon cavity, two extra side-peaks at $eV_{\rm pg}^{\rm GS;FES} = eV_{\rm pg}^{\rm GM;FEM} -
\hbar\omega_{\rm ph}$  are observed in addition 
to the original main-peaks at $V_{\rm pg}^{\rm GM;FEM}$.
The superscripts $\rm GM$($\rm FEM$) refers to the ground states(first-excited state) main-peak, respectively,
and $\rm GS$($\rm FES$) stands for photon-induced ground state(first-excited state) side-peak, respectively.

The side-peaks indicate the PAT, where the system
satisfies $e|V_{\rm pg}^{\rm GM;FEM}-V_{\rm pg}^{\rm GS;FES}|
\cong \hbar\omega_{\rm ph} $.\cite{Kouwenhoven73.3443}
The two new side-peaks at $V_{\rm pg}^{\rm GS} = 0.95$~mV and $V_{\rm pg}^{\rm FES} = 0.55$~mV shown in \fig{I-Xp} 
are caused by photon-replica of the ground state and photon-replica of the first-excited state,
respectively. We find that the separation of the photon replica side-peaks from the original
main-peaks corresponds to the photon energy. 

It should be noted that the current in the photon-induced side-peaks is strongly enhanced 
by increasing the electron-photon coupling strength. 
Thus the photon-induced side-peaks
exhibits a PAT process with different photon absorption mechanism
from the main-peaks.

\begin{figure}[htbq]
 \includegraphics[width=0.23\textwidth,angle=0]{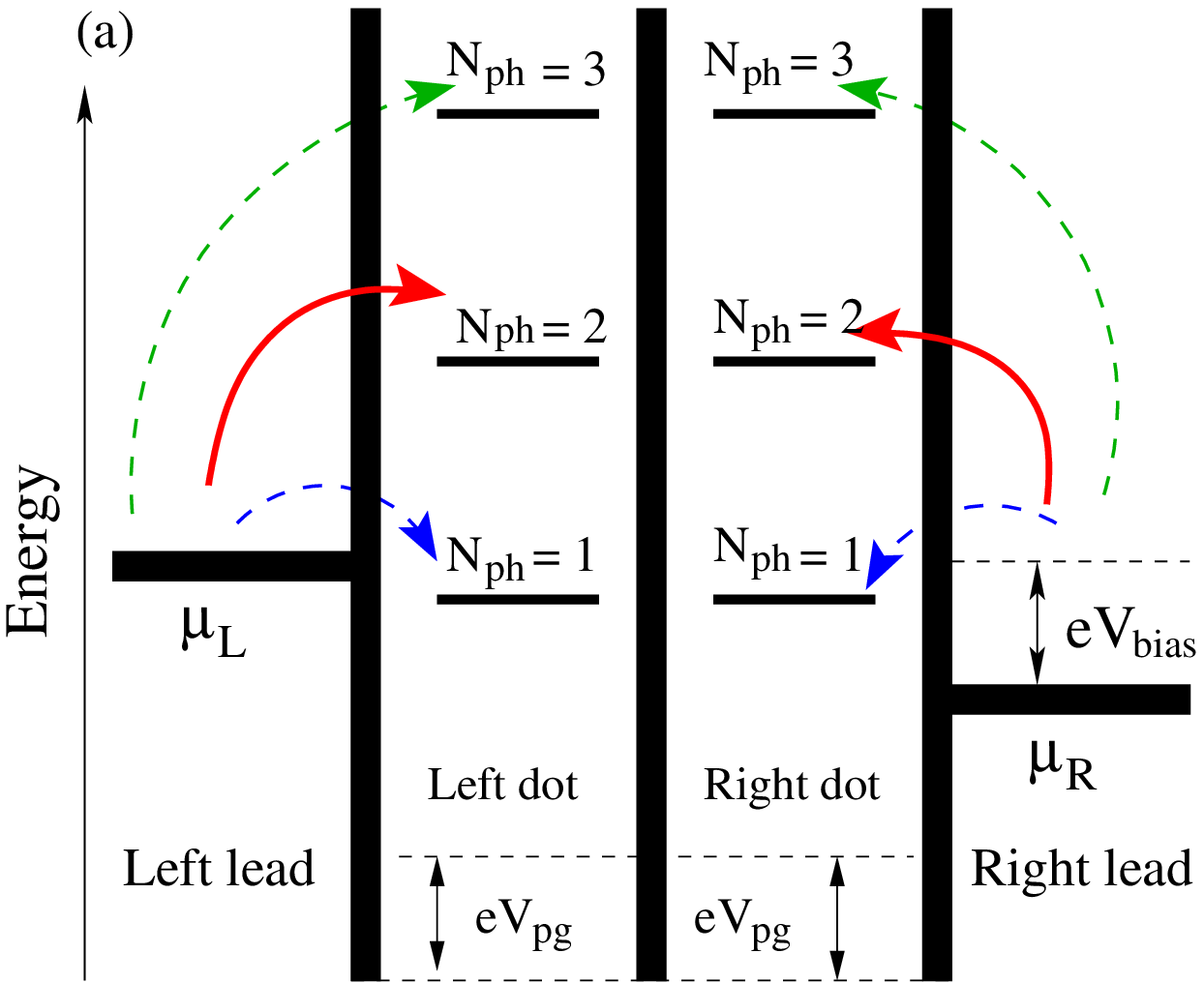}
 \includegraphics[width=0.23\textwidth,angle=0]{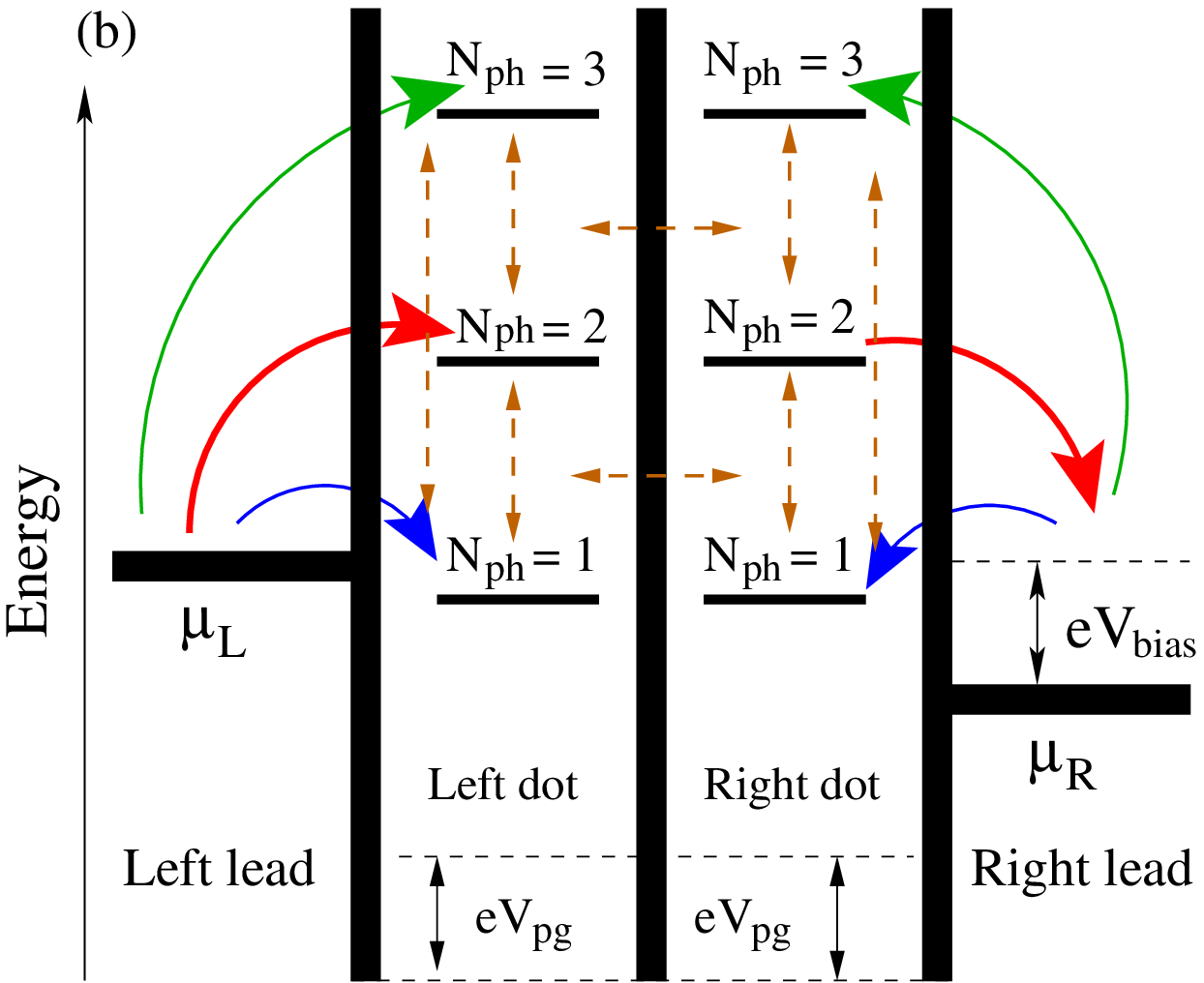}
\caption{(Color online) Schematic representation of photon-activated resonance energy
levels and electron transition by changing the plunger gate voltage $V_{\rm pg}$ in the
photon-induced first-excited state side-peak at $V_{\rm pg}^{\rm FES} = 0.55$~mV (a), 
and the photon-induced ground state side-peak at $V_{\rm pg}^{\rm GS} = 0.95$~mV(b) of the \fig{I-Xp}. 
The DQD-system is embedded in a photo cavity with the photon energy $\hbar \omega_{\rm ph}$ and photon content
$N_{\rm ph}$ in each many-body state.  The chemical potential difference is $e V_{\rm bias} = \Delta \mu
= \mu_L - \mu_R$.} \label{E_Level}
\end{figure}

In order to show the dynamics of the PAT process involved in the formation of the photon-induced side-peaks
in the left and right current shown in figure \ref{I-Xp}, 
we schematically present the photon absorption process in \fig{E_Level}.

Figure \ref{E_Level}(a) demonstrates the tunneling processes forming the FES at $V_{\rm pg}^{\rm FES} = 0.55$~mV. 
The electron from the left or the right lead absorbs two photons and is transferred to the MB states 
containing two photons $N_{\rm ph} = 2$ (red solid arrows) situated above the 
bias window with photon energy $\hbar\omega_{\rm ph}$.
The electron tunneling process in the states containing one photon $N_{\rm ph} = 1$ (blue dashed arrows) 
and three photons $N_{\rm ph} = 3$ (green dashed arrows) are very weak.
Figure \ref{E_Level}(b) shows the dynamical mechanism that makes the GS at $V_{\rm pg}^{\rm GS} = 0.95$~mV. 
In addition to the electron tunneling in the state containing two photons $N_{\rm ph} = 2$ (red solid arrows), 
the tunneling process in one photon state $N_{\rm ph} = 1$ (blue solid arrows)
and three photons $N_{\rm ph} = 3$ states (green solid arrows) become active.
The tunneling mechanism here is a multiphoton absorption process with up to three photons with 
a strong inter-dot tunneling. In which the electron is scattered 
between one, two, and three photon(s) states by absorbing and emitting photon energy 
$N_{\rm ph}\times\hbar\omega_{\rm ph}$. 
The two photons state here has a shorter lifetime than the two photons states in the FES, because whenever an electron
from the left lead tunnels into the two photon states in the DQD system it subsequently directly tunnels out to the right lead.

\begin{figure}[htbq]
      \includegraphics[width=0.23\textwidth]{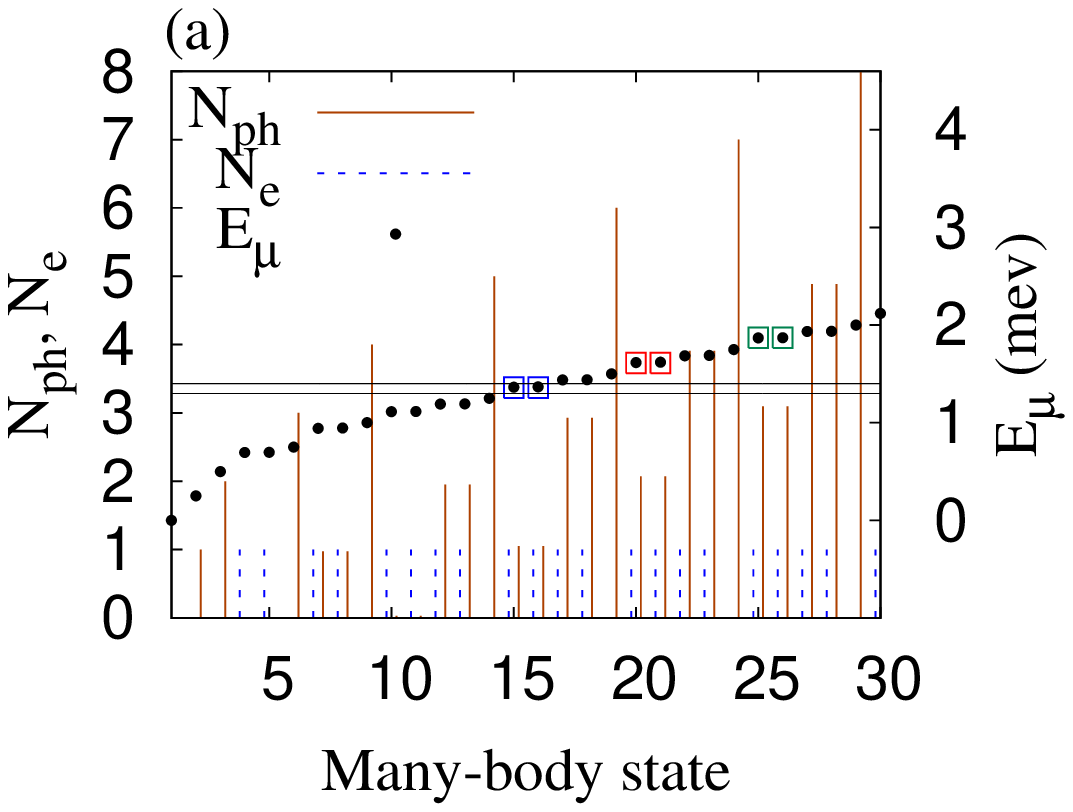}
      \includegraphics[width=0.23\textwidth]{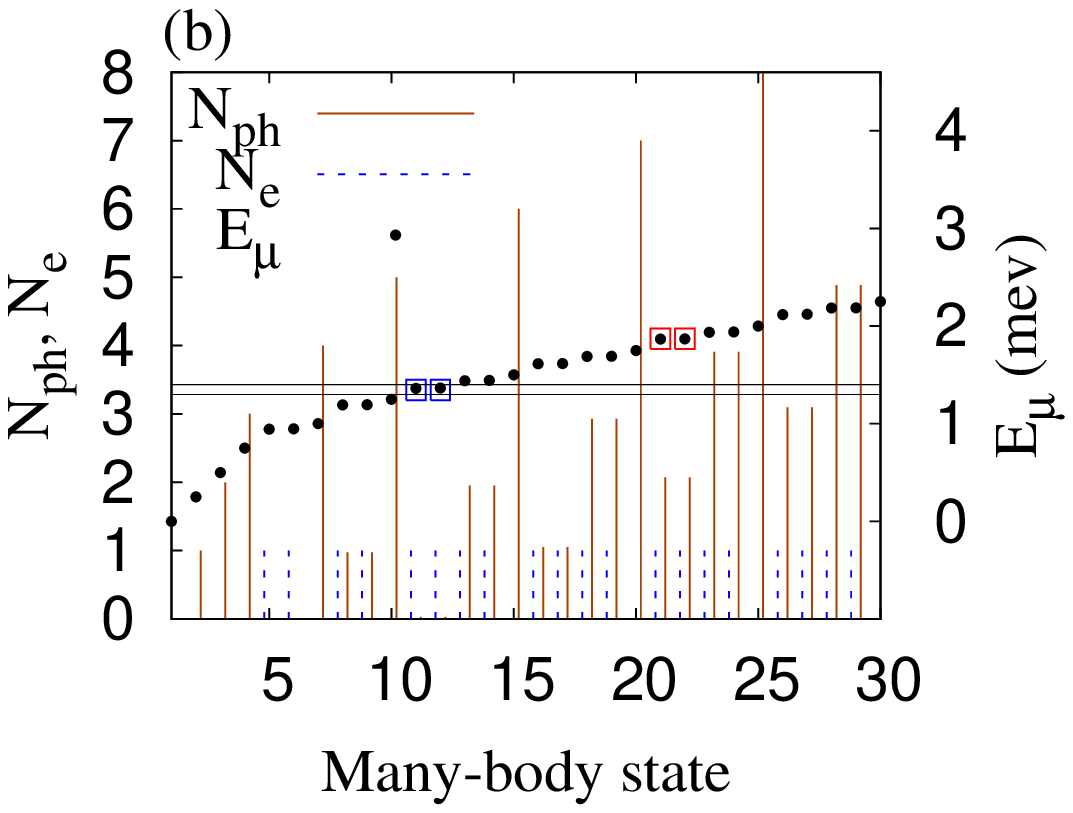}\\
      \includegraphics[width=0.23\textwidth]{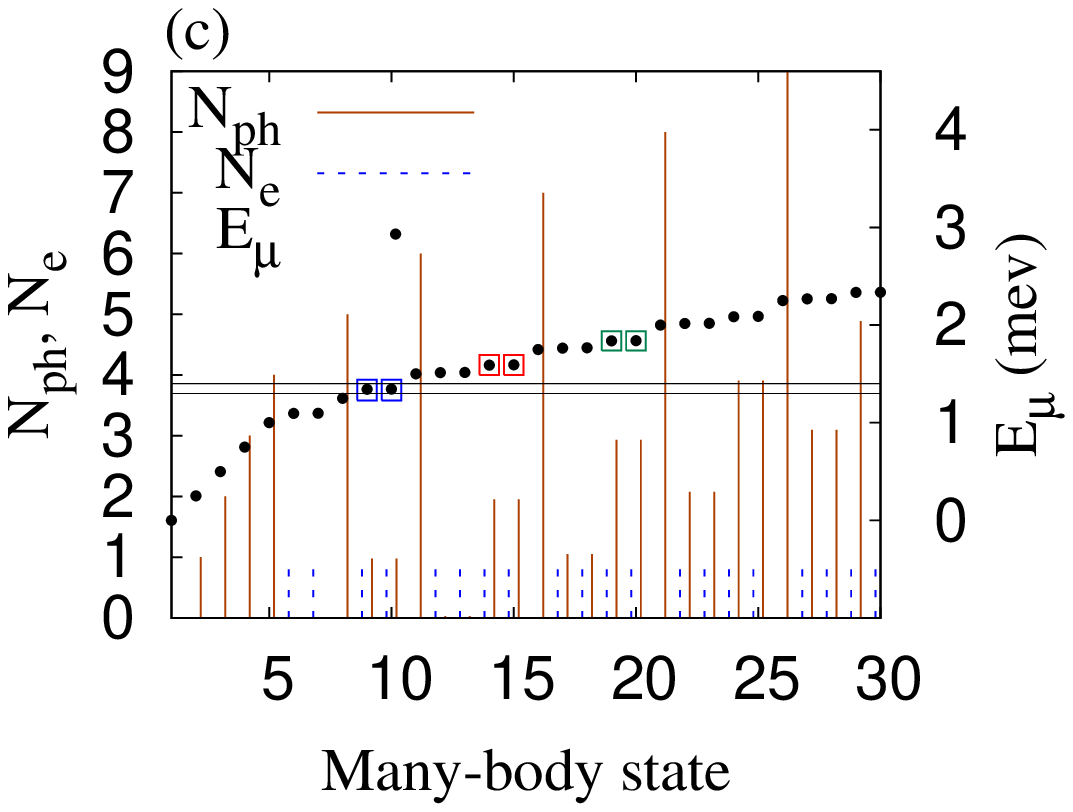}
      \includegraphics[width=0.23\textwidth]{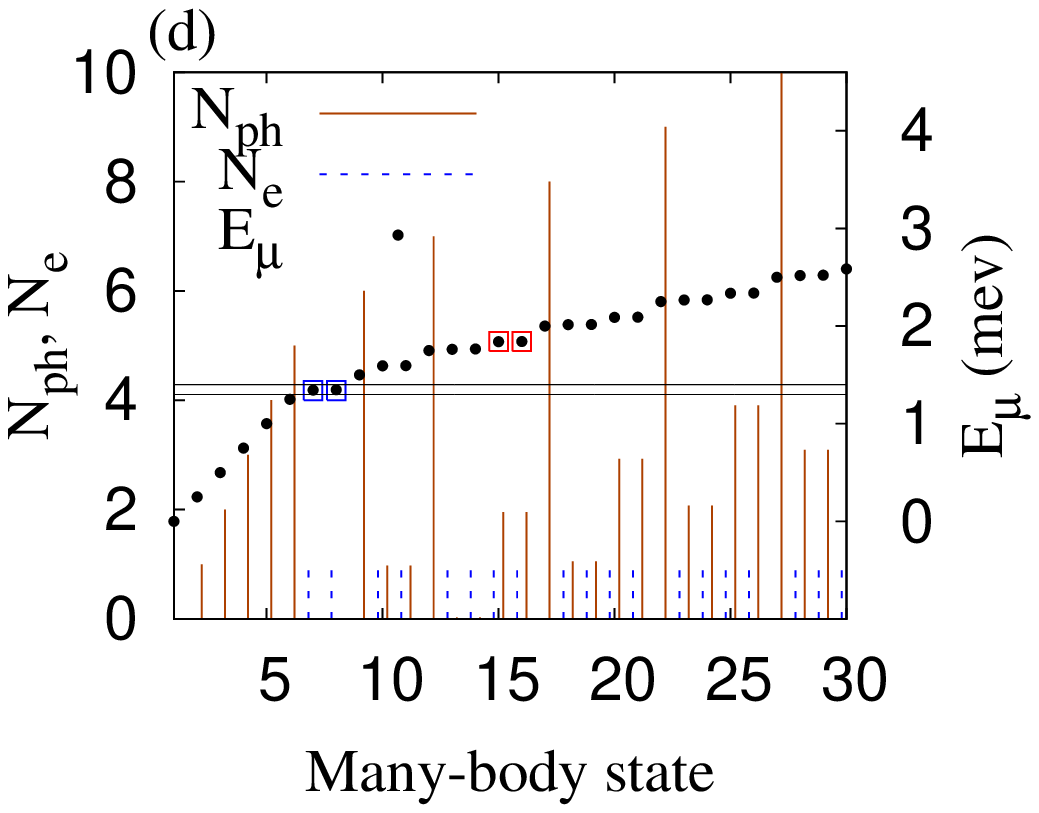}
      \caption{(Color online) The MB energy spectrum $E_{\mu}$ (dotted black), the mean
               electron number in the MB state $|\breve{\mu})$ (blue dashed line),
               the mean photon number $N_{\rm ph}$ (red line) in the case of
               $x$-polarized field with plunger-gate voltage 
               at (a) photon-induced first-excited side-peak $V_{\rm pg}^{\rm FES} = 0.55$~mV, 
                  (b) first-excited main-peak $V_{\rm pg}^{\rm FEM} = 0.8$~mV, 
                  (c) photon-induced ground-state side-peak $V_{\rm pg}^{\rm GS} = 0.95$~mV, and 
                  (d) ground-state main-peak $V_{\rm pg}^{\rm GM} = 1.2$~mV of \fig{I-Xp}
                  for the case of $g_{\rm ph} = 0.1$~meV (blue solid line).
                  The chemical potentials are $\mu_{\rm L} = 1.4$~meV and $\mu_{\rm L} = 1.3$~meV (black line),
                  thus $\Delta \mu =0.1~{\rm meV}$.
                 Other parameters are $B = 0.1~{\rm T}$, and $\hbar \omega_{\rm ph} = 0.25$~meV.
                 The color of the square is referred to in the text.}
\label{NeNphSz_Xp}
\end{figure}

To further illustrate the characteristics of the most active MB states in the tunneling process 
forming the two main current peaks and the two side current peaks in the \fig{I-Xp}, we present
\fig{NeNphSz_Xp} which shows the characteristics of the MB states at plunger-gate voltage $V_{\rm pg}^{\rm FES} = 0.55$~mV (a),
$V_{\rm pg}^{\rm FEM} = 0.8$~mV (b), $V_{\rm pg}^{\rm GS} = 0.95$~mV (c), and $V_{\rm pg}^{\rm GM} = 1.2$~mV (d) in
the case of $g_{\rm ph} = 0.1$~meV. 

Figure \ref{NeNphSz_Xp}(a) indicates how the FES is contributed to by the MB states at $V_{\rm pg}^{\rm FES} = 0.55$~mV.
Since there are two photons initially in the cavity, we shall seek the MB-states that contain one, two, and three photon(s),
to observe multiple inelastic electron scattering in the states of DQD system at the side-peaks.\cite{Kouwenhoven.75.2003}
Here, we focus on six MB states, two inactive MB states 
$|\breve{15})$ and $|\breve{16})$ in the bias window (blue squared dot) 
with $N_{\rm ph} = 1.052$ in each state and the energies $E_{15} = 1.364$~meV and $E_{16} = 1.366$~meV.
There are four MB state above the bias window: Two photon-activated states 
$|\breve{20})$ and $|\breve{21})$ (red squared dot) with $N_{\rm ph} = 2.073$ in each state and energies
$E_{20} = 1.616$~meV and $E_{21} = 1.618$~meV, and two more MB states $|\breve{25})$ and $|\breve{26})$ (green squared dot)
with $N_{\rm ph} = 3.094$ in each state and energies $E_{25} = 1.867$~meV and $E_{26} = 1.870$~meV.
We clearly see that the energy difference between the inactive states and the photon-activated states is appropriately
equal to the $(N_{\rm ph,ac}-N_{\rm ph,in})\times \hbar\omega_{\rm ph}$, where
$N_{\rm ph,ac}$ and $N_{\rm ph,in}$ are the photon number in the photon-activated states and the inactive
states, respectively.
We observe that the electrons can undergo the following possible tunneling process: 
An electron from either lead may absorb two photons from the cavity
being transferred to two photons states $|\breve{20})$ and $|\breve{21})$ with absorption energy 
$E_{20}- E_{15} = (N_{\rm ph,20} - N_{\rm ph,15})\times \hbar\omega_{\rm ph} \simeq 0.252$~meV 
or $E_{21}- E_{16} = (N_{\rm ph,21} - N_{\rm ph,16})\times \hbar\omega_{\rm ph} \simeq 0.252$~meV 
which is approximately equal to the energy required to transfer an electron from the leads to two photons states 
as schematically shown in \fig{E_Level}(a). Therefore, the two photon absorption mechanism dominates here 
without making electron inelastic scattering to the one and three photon states.
The electron tunneling process from the leads to the DQD system suggest that the electrons are collected in either 
individual dot.

Figure \ref{NeNphSz_Xp}(b) shows the MB states of the first-excited state main-peak at $V_{\rm pg}^{\rm FEM} = 0.8$~mV. 
There are two inactive states $|\breve{11})$ and $|\breve{12})$ (blue squared dot) with energies $E_{11} =
1.362$~meV and $E_{12} = 1.364$~meV in the bias window ($N_e = 1$, $N_{\rm ph} = 0.029$) and two photon-activated
states $|\breve{21})$, and $|\breve{22})$ (red squared dot) with energies $E_{21} = 1.866$~meV and $E_{22} =
1.868$~meV above the bias window ($N_e = 1$, $N_{\rm ph} = 2.073$).
The photon-activated states that contain two photons are responsible for the electron transport
with energy values $E_{21} - E_{11} \cong (N_{\rm ph,21}-N_{\rm ph,11})\times \hbar\omega_{\rm ph} \cong 0.504$~meV 
or $E_{22} - E_{12} \cong (N_{\rm ph,22}-N_{\rm ph,12})\times \hbar\omega_{\rm ph} \cong 0.504$~meV.

Figure \ref{NeNphSz_Xp}(c) demonstrates the MB states that participate 
to the electron transport in the GS at $V_{\rm pg}^{\rm GS} = 0.95$~mV.
The electron transport mechanism here is different from the one for the FES.
The contributions to the GS is by the following significant MB states: Two active MB states 
$|\breve{9})$ and $|\breve{10})$ (blue squared dot) 
containing $N_{\rm ph} = 0.976$ with energies $E_{9} = 1.342$~meV and $E_{10} = 1.344$~meV are
located in the bias window, two photon-activated states $|\breve{14})$ and $|\breve{15})$ (red squared dot)
have $N_{\rm ph} = 1.95$ above the bias window with energies $E_{14} = 1.5901$~meV and $E_{15} = 1.5902$~meV, 
and two more photon-activated states $|\breve{19})$ and $|\breve{20})$ (green squared dot) contain
$N_{\rm ph} = 2.93$ with energies $E_{19} = 1.838$~meV and $E_{20} = 1.840$~meV.
Significantly, these six MB states participate in the electron transport with the following important photon 
absorption processes with inter-dot tunneling as schematically shown previously in \fig{E_Level}(b): 
(1) Electron from either lead absorbs one photon tunneling to the one photon states $|\breve{9})$ and $|\breve{10})$,
(2) An electron from the left lead absorbs two photons and is transferred to two photons states
$|\breve{14})$ and $|\breve{15})$ with absorption energy 
$E_{14}- E_{9} = (N_{\rm ph,14} - N_{\rm ph,9})\times \hbar\omega_{\rm ph} \simeq 0.248$~meV 
or $E_{15}- E_{10} = (N_{\rm ph,15} - N_{\rm ph,10})\times \hbar\omega_{\rm ph} \simeq 0.247$~meV
which is approximately 
equal to the energy required to transfer electron from one photon state to two photons states, then the electron 
tunnels to the right lead emitting photons,
(3) Absorbing three photons, an electrons from either the left lead or the right lead transfers to three photons states 
$|\breve{19})$ and $|\breve{20})$ with energy
$E_{19}- E_{9} = (N_{\rm ph,19} - N_{\rm ph,9})\times \hbar\omega_{\rm ph} \simeq 0.496$~meV
or $E_{20}- E_{10} = (N_{\rm ph,20} - N_{\rm ph,10})\times \hbar\omega_{\rm ph} \simeq 0.496$~meV
that is approximately the energy amount needed to transfer an electron to a three photons state.
The tunneling processes from the leads to the DQD system and all activated six-MB states
suggest that the electrons perform multiple scattering absorption and emission processes between 
the states in each individual dot with inter-dot tunneling.
These possible tunneling processes indicate to us that the existence of the FES is caused by 
multiphoton absorption processes with up to three photons. 
In addition, we should mention that the tunneling process from the left lead 
to two photon states in the DQD system and the tunnel out to the right lead decreases
the dwell time of electron in the central system, while the dwell time of electron in the FES
was longer due to charge accumulation in the DQD system.

Figure \ref{NeNphSz_Xp}(d) demonstrates the MB states of the ground state main-peak at $V_{\rm pg}^{\rm GM} = 1.2$~mV. Four 
MB states are also important here, two inactive states $|\breve{7})$ and $|\breve{8})$ (blue squared dot) with energies $E_{7} =
1.344$~meV and $E_{8} = 1.346$~meV in the bias window ($N_e = 1$, $N_{\rm ph} = 0.029$) and two photon-activated
states $|\breve{15})$ and $|\breve{16})$ (red squared dot) with energies $E_{15} = 1.840$~meV and $E_{16} =
1.842$~meV above the bias window ($N_e = 1$, $N_{\rm ph} = 1.951$). The energy difference between
photon-activated states above the bias window and the inactive states in the bias window satisfies 
the same rule of the FEM, such that
$E_{15} - E_{7} \cong (N_{\rm ph,15}-N_{\rm ph,7})\times \hbar\omega_{\rm ph} \cong 0.496$~meV 
and $E_{16} - E_{8} \cong (N_{\rm ph,16}-N_{\rm ph,8})\times \hbar\omega_{\rm ph} \cong 0.496$~meV.

These results suggest that each photon-activated state above the bias window has 
two more photons than the inactive states in the bias window at both the main peaks.
When an electron from the left lead tunnels to the DQD system it absorbs (or forms a quasi-particle with) 
two photons from the cavity and is transferred
to the photon-activated states above the bias window, then it tunnels to the right lead.
Therefore, both main-peaks are caused by two photons absorption processes in the transport.

Observing all these photon activated processes it is important to have in mind that the fact that
we retained the dia- and the paramagnetic parts of the electron-photon interaction (\eq{H_Int}) 
thus allowing for a broader range of transitions possibilities than only the paramagnetic term
describes. 

\begin{figure}[htbq]
  \includegraphics[width=0.23\textwidth]{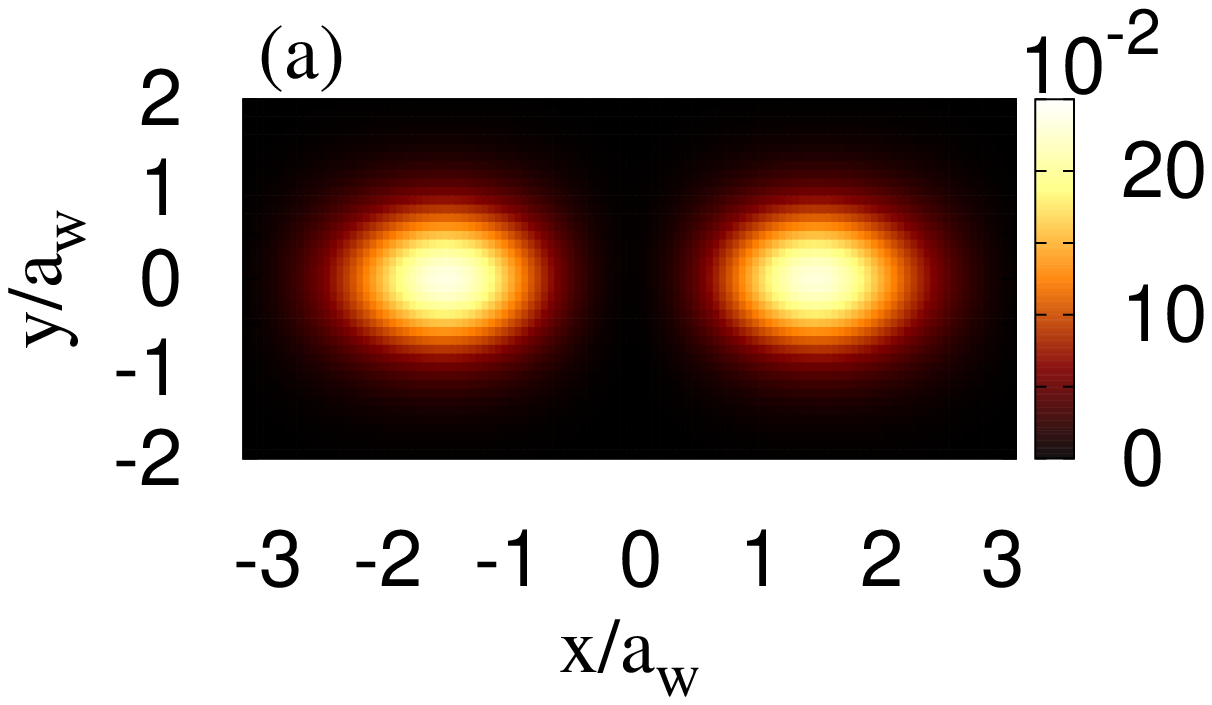}
 \includegraphics[width=0.23\textwidth]{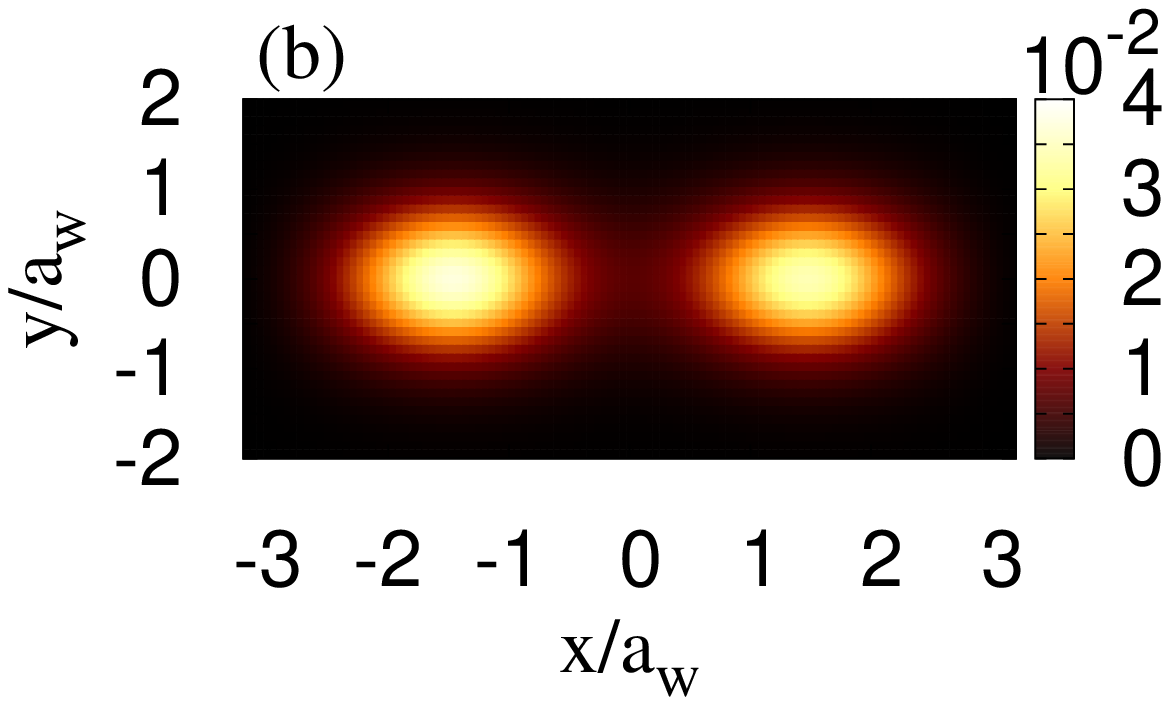}\\
 \includegraphics[width=0.23\textwidth]{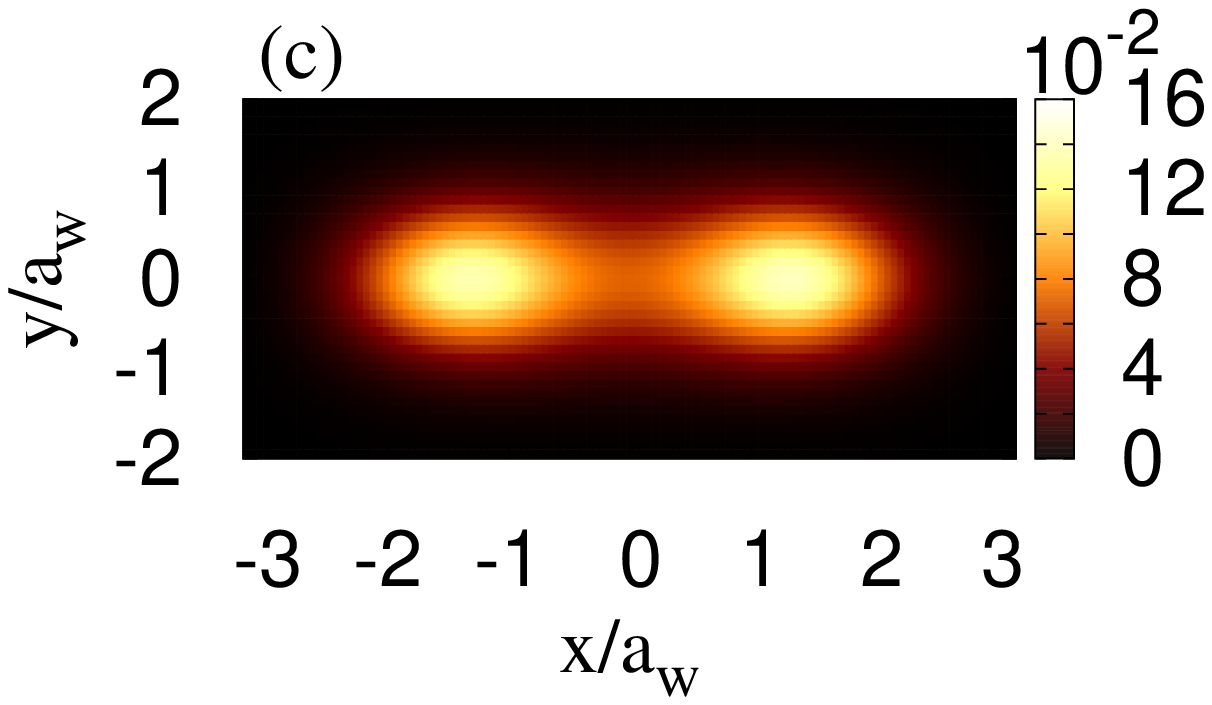}
 \includegraphics[width=0.23\textwidth]{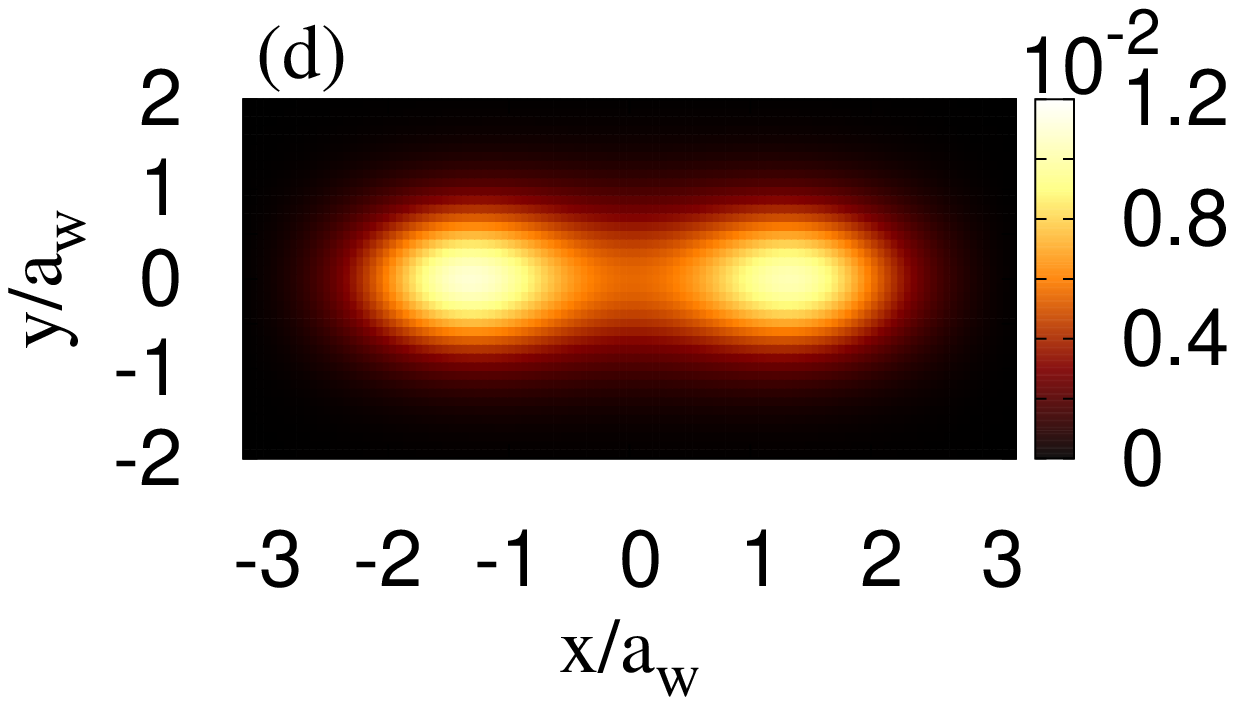}\\
\caption{(Color online) The charge density distribution of
the DQD system with $x$-polarized photon field at time $220$~ps corresponding to
  (a) photon-induced first-excited side-peak $V_{\rm pg}^{\rm FES} = 0.55$~mV, 
  (b) first-excited main-peak $V_{\rm pg}^{\rm FEM} = 0.8$~mV, 
  (c) photon-induced ground-state side-peak $V_{\rm pg}^{\rm GS} = 0.95$~mV, and 
  (d) ground-state main-peak $V_{\rm pg}^{\rm GM} = 1.2$~mV of \fig{I-Xp}.
for the case of $g_{\rm ph} = 0.1$~meV (blue solid line). Other parameters are $\hbar \omega_{\rm ph} =
0.25~{\rm meV}$, $B=0.1$~T, $a_{w} = 23.8~{\rm nm}$, $L_x = 300~{\rm nm}$, and $\hbar
\Omega_0 = 2.0~{\rm meV}$.} \label{Q_Xp}
\end{figure}

Figure \ref{Q_Xp} indicates the charge density distribution in the DQD system with $x$-polarized photon field
at plunger-gate $V_{\rm pg}^{\rm FES} = 0.55$~mV (a), $V_{\rm pg}^{\rm FEM} = 0.8$~mV (b), 
$V_{\rm pg}^{\rm GS} = 0.95$~mV (c), and $V_{\rm pg}^{\rm GM} = 1.2$~mV (d) of the \fig{I-Xp}
at time $t = 220$~ps and $g_{\rm ph} = 0.1$~meV.
In the FES at $V_{\rm pg}^{\rm FES} = 0.55$~mV, the electron charge density forms
two peaks which are strongly localized in the dots without inter-dot tunneling shown in \fig{Q_Xp}(a), thus  
the electron dwell time is increased and the electrons stay longer time in the DQD system.

In the case of $V_{\rm pg}^{\rm FEM} = 0.8$~mV, the electrons are accumulated in the dots with
a weak inter-dot tunneling. Comparing to the case with no photon cavity \fig{Q_e-e}(a), a slight
inter-dot tunneling is observed indicating charge polarization between the dots. As a result,
the electron charge density is slightly enhanced in the $x$-polarized photon field, but more 
importantly it is also slightly delocalized resulting in a higher conductance through the serial dot molecule.

In the GS at $V_{\rm pg}^{\rm GS} = 0.95$~mV, the electron charge density is enhanced
and exhibits charge accumulation in the dots with a very strong inter-dot tunneling shown in \fig{Q_Xp}(c)
which decreases the electron dwell time in the DQD-system.
This is the reason why the current in the GS is relatively higher than the current is FES.

In the case of $V_{\rm pg}^{\rm GM} = 1.2$~mV, the electron-photon interactions does not
have a big effect on the charge density distribution, the charge distribution of the dots is
already overlapping.

\subsection{$y$-photon polarization  (TE$_{101}$ mode)}

In this section, we assume the photon-cavity is linearly polarized in the $y$-direction
with photon energy $\hbar \omega_{\rm ph} = 0.25$~meV and initially two photons in the single photon mode.
The MB energy spectrum of the DQD-system in the $y$-polarization is
very similar to that in the $x$-polarization photon mode as shown in \fig{MBE_Xp}. Two extra MB states are observed with
the spin states of the ground state and first-excited state in the bias window, the extra MB states indicate
the photon-replica states in the presence of the photon cavity.

\begin{figure}[tbhq]
 \includegraphics[width=0.5\textwidth]{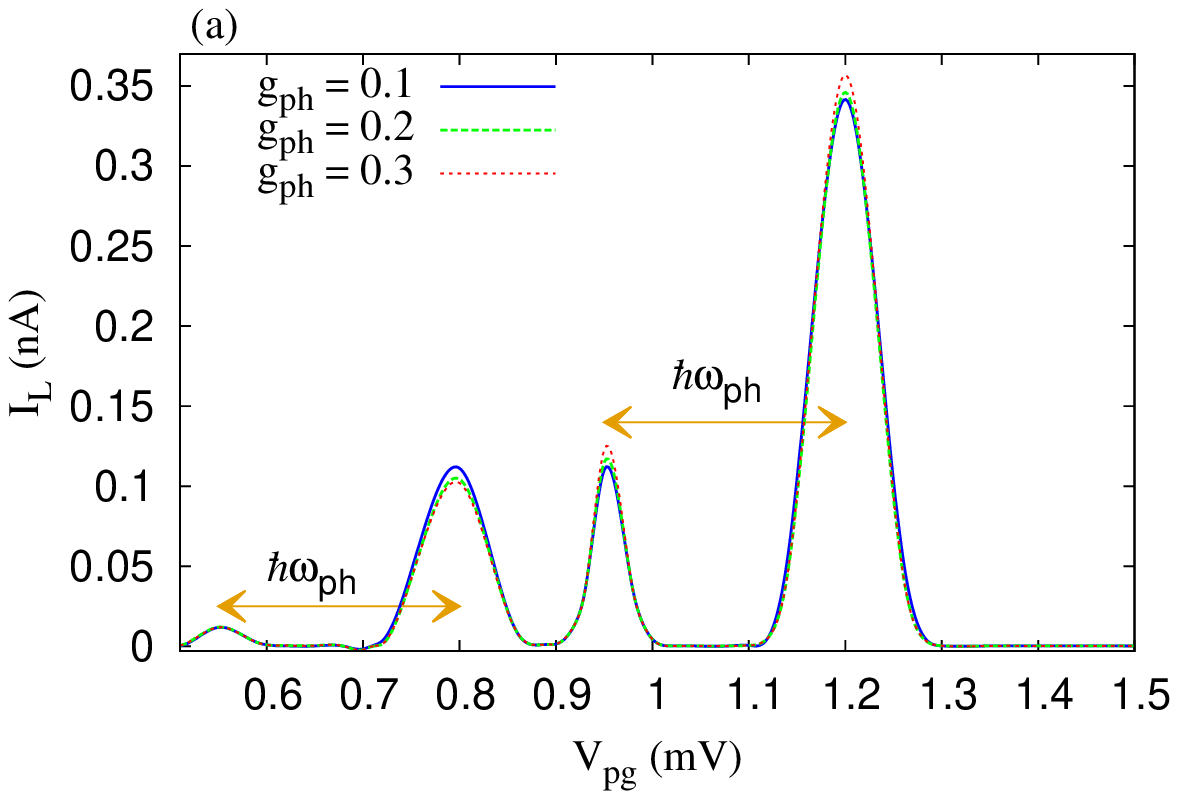}\\
 \includegraphics[width=0.5\textwidth]{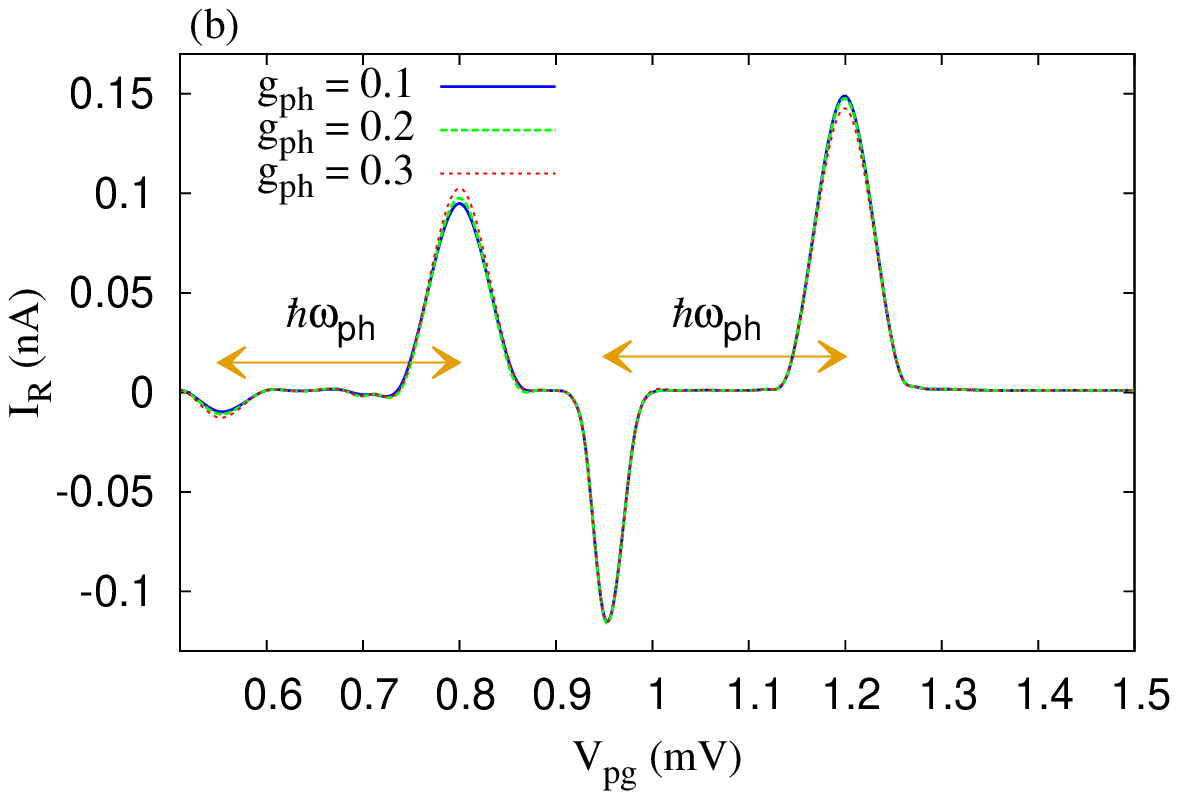}
\caption{(Color online) The left current (a) and the right current (b) 
versus the plunger gate voltage $V_{\rm pg}$ at time ($t = 220~{\rm ps}$) 
in the case of $y$-polarized photon field.  The electron-photon coupling is changed to 
be $g_{\rm ph} = 0.1$~meV (blue solid), $0.2$~meV (green dashed), 
and $0.3$~meV  (red dotted).
Other parameters are $\hbar\omega_{\rm ph} = 0.25~{\rm meV}$, $\Delta \mu =0.1~{\rm meV}$,
and $B = 0.1~{\rm T}$.}
 \label{I-Yp}
\end{figure}

Figure \ref{I-Yp} shows the left current (a) and the right current (b) at time $t = 220$~ps
for different electron-photon coupling strength  
$g_{\rm ph} = 0.1~{\rm meV}$, (blue solid), $0.2~{\rm meV}$ (green dashed), 
and $0.3~{\rm meV}$ (red dotted).
Similar to the $x$-polarized photon field, two extra photon-induced side-peaks at $V_{\rm pg}^{\rm FES} = 0.55$~mV
and $V_{\rm pg}^{\rm GS} = 0.95$~mV are observed with the main-peaks at $V_{\rm pg}^{\rm FEM} = 0.8$~mV
$V_{\rm pg}^{\rm GM} = 1.2$~mV. A very weak current enhancement in the photon-induced side-peaks is predicted by increasing 
the electron-photon coupling strength, while the current enhancement in the photon-induced side-peaks 
is very strong in the $x$-polarized photon field shown in \fig{I-Xp}. The weaker effects for the $y$-polarization
are expected since the photon energy is farther from resonance for states describing motion in that direction, i.e.\
the confinement energy in the $y$-direction is much higher.

The characteristics of the MB states in the bias window and above the bias window
in the $y$-polarized photon field are very similar to that in the $x$-direction shown in \fig{NeNphSz_Xp}. 
The main-peaks and the photon-induced side-peaks in the $y$-polarized photon are contributed to by almost the same
absorption processes of the $x$-polarized photon.

\begin{figure}[htbq]
 \includegraphics[width=0.23\textwidth]{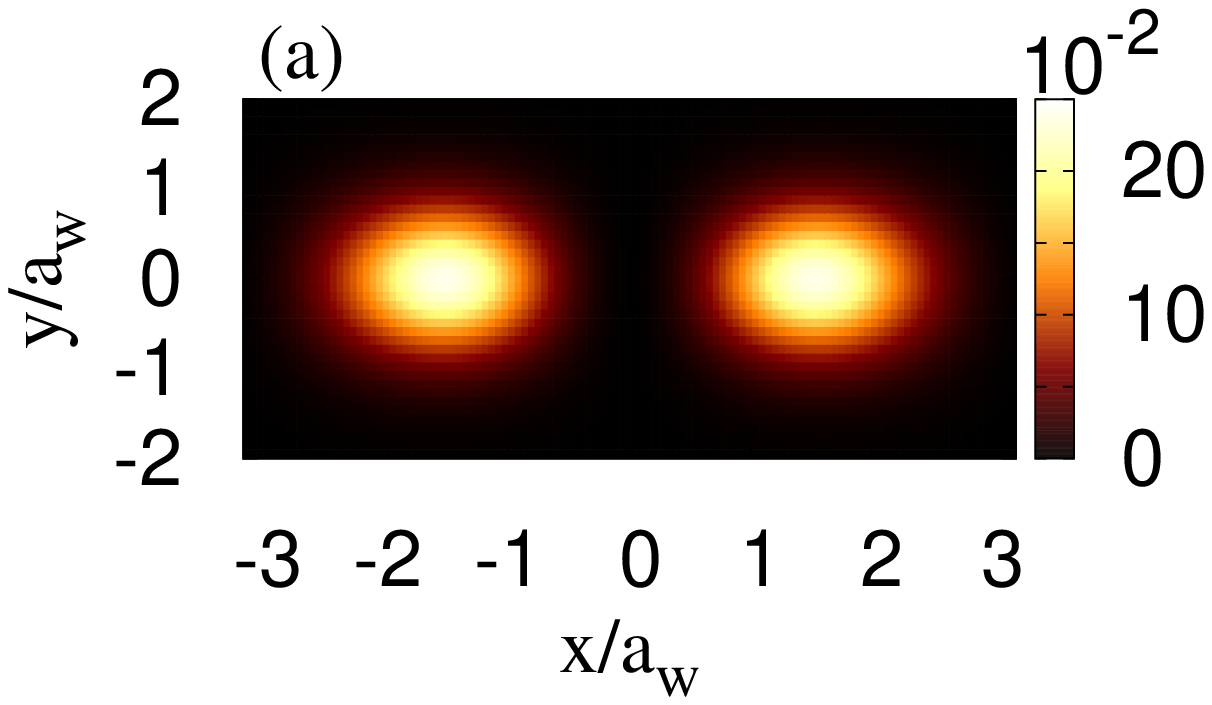}
 \includegraphics[width=0.23\textwidth]{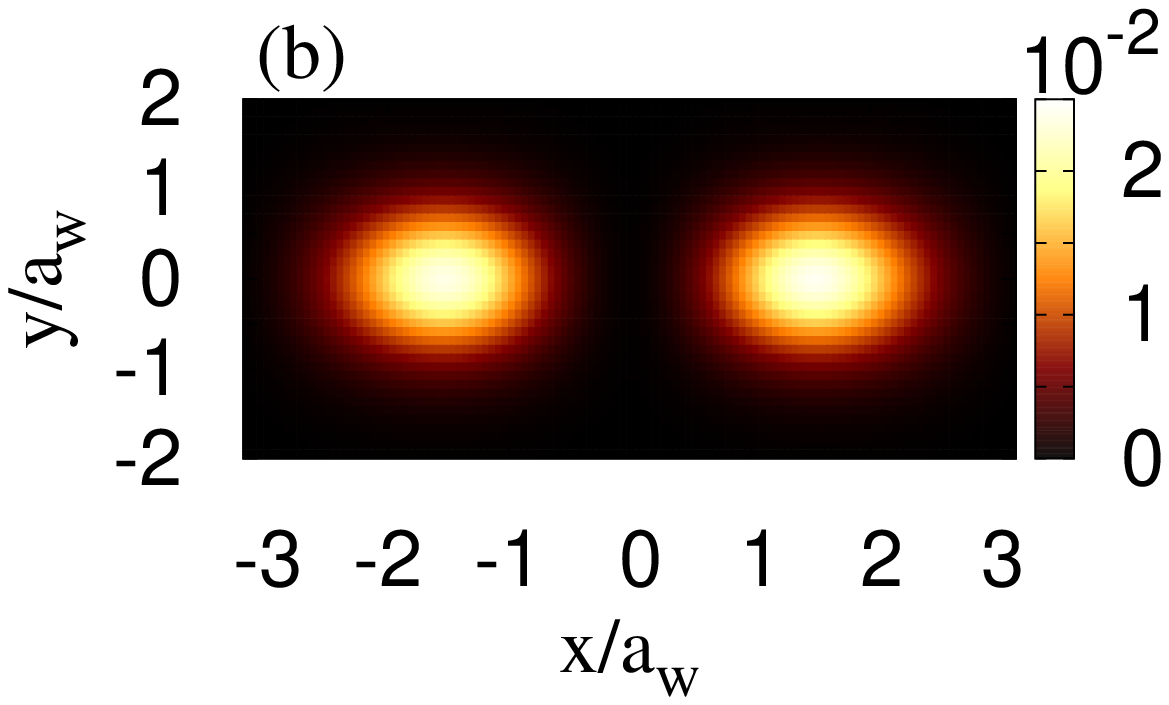}\\
 \includegraphics[width=0.23\textwidth]{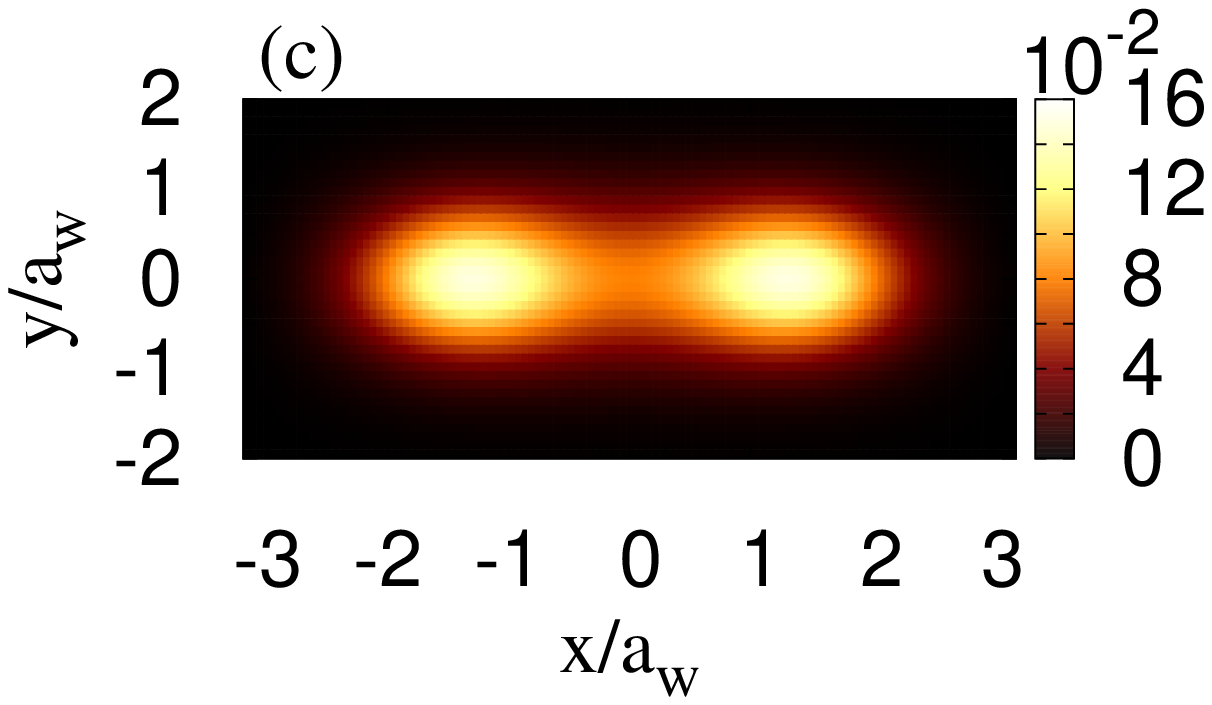}
 \includegraphics[width=0.23\textwidth]{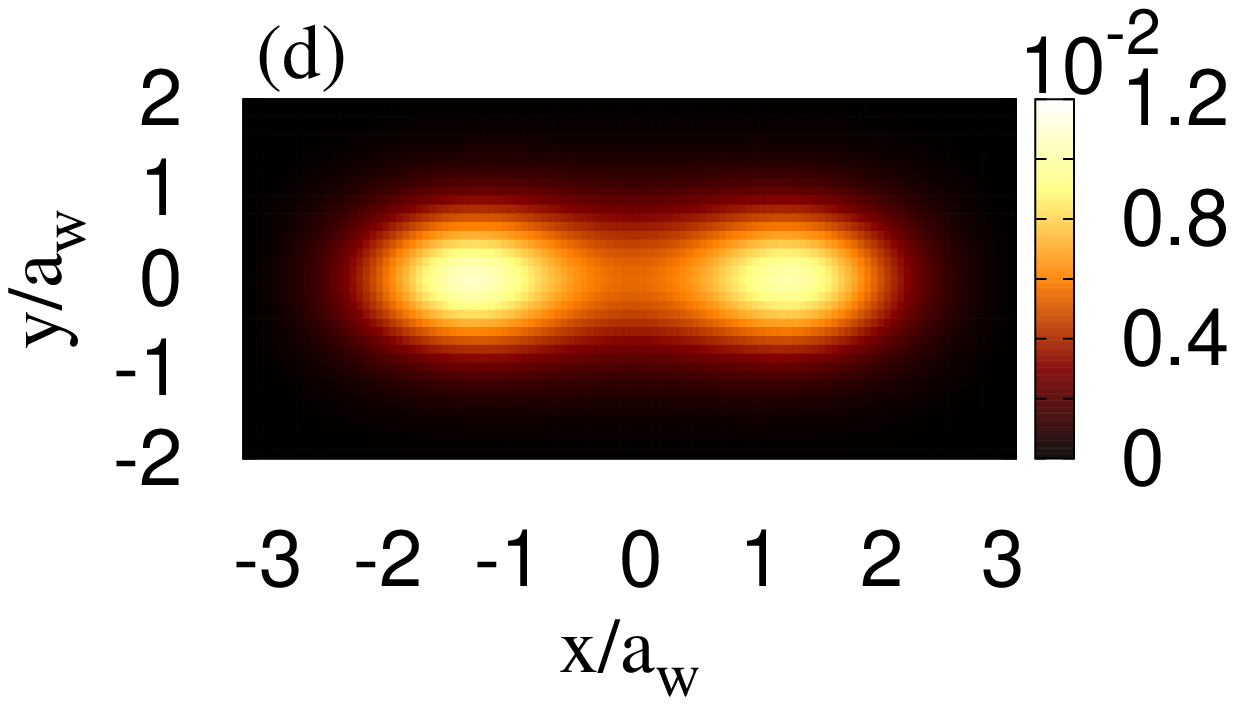}\\
\caption{(Color online) The charge density distribution of
the DQD system with $y$-polarized photon field at time $220$~ps corresponding to
(a) photon-induced first-excited side-peak $V_{\rm pg}^{\rm FES} = 0.55$~mV, 
(b) first-excited main-peak $V_{\rm pg}^{\rm FEM} = 0.8$~mV, 
(c) photon-induced ground-state side-peak $V_{\rm pg}^{\rm GS} = 0.95$~mV, and 
(d) ground-state main-peak $V_{\rm pg}^{\rm GM} = 1.2$~mV of \fig{I-Yp}
for the case of $g_{\rm ph} = 0.1$~meV (blue solid line). Other parameters are $\hbar \omega_{\rm ph} =
0.25~{\rm meV}$, $B=0.1$~T, $a_{w} = 23.8~{\rm nm}$, $L_x = 300~{\rm nm}$, and $\hbar
\Omega_0 = 2.0~{\rm meV}$.} \label{Q_Yp}
\end{figure}

Figure \ref{Q_Yp} demonstrates the charge density distribution in the DQD system in the case of $y$-polarized photon field
at  plunger-gate voltage $V_{\rm pg}^{\rm FES} = 0.55$~mV FES (a), $V_{\rm pg}^{\rm FEM} = 0.8$~mV FEM (b),
$V_{\rm pg}^{\rm GS} = 0.95$~mV GS (c), and $V_{\rm pg}^{\rm GM} = 1.2$~mV GM (d) shown in the \fig{I-Yp},  at time $t = 220$~ps and
$g_{\rm ph} = 0.1$~meV.
In the case of $V_{\rm pg}^{\rm FES} = 0.55$~mV (FES), the electrons are strongly
localized in the dots with no electron tunneling from the left-dot to the right-dot.

In FEM at $V_{\rm pg}^{\rm FEM} = 0.8$~mV, the electron makes a resonance state localized
in each dot without inter-dot tunneling, while a weak inter-dot tunneling was observed
in the $x$-polarized photon field at FEM shown in \fig{Q_Xp}(b). Therefore, the electron dwell time in the DQD system 
in $y$-polarized photon is longer than that in the $x$-polarized photon at FEM.

But at $V_{\rm pg}^{\rm GS} = 0.95$~mV (GS), the inter-dot tunneling is very strong and 
electrons prefer to make inelastic multiple scattering in each dot with inter-dot tunneling.

In GM at $V_{\rm pg}^{\rm GM} = 1.2$~mV, the electrons form a state accumulated in the dots
with a strong electron tunneling between the dots similar to the charge density distribution in
the $x$-polarization shown in \fig{Q_Xp}(d). Thus, the current in the GM is higher than that
in the FES.

\section{Conclusions}\label{Sec:IV}

Photon-assisted transient electron transport through a DQD system placed
in a photon cavity with initially two linearly polarized photons 
can be controlled by a plunger-gate voltage. 
The serial double quantum dot molecule is important since the two lowest
states of it have very different properties reflected in the fact
that the wavefunction of one is symmetric, but antisymmetric of the other.

We analyzed the electron transport through the system without 
and with a photon cavity by using a non-Markovian QME formalism.
In the absence of a photon cavity, two current peaks were found: Ground state peak and
a peak due to the first-excited state, originating from resonance energy levels of the DQD system with the
first-subband energy of the leads. 
These two states could be used for a qubit in a quantum computer, in which 
the ground state resonance exhibits a strong inter-dot electron tunneling 
while the electrons in the first-excited state resonance form a state localized in each dot.

In the presence of either longitudinally or transversely polarized cavity photon field,
two extra side-peaks are found: A peak due to the photon-replica of the ground state, and 
a photon replica of the first-excited state. The appearance of side-peaks is due to PAT
of electrons in the DQD system.
The characteristics of the photon activated MB states have been used to analyze
the nature of the PAT. The peak due to the photon-replica of ground state is
caused by multiphoton absorption processes with up to three photons. 
In the peak caused by the photon replica of the first-excited state, 
the electrons in the leads are transferred to two-photons states
in the DQD system accumulation charge in the individual dots without inter-dot tunneling.
Furthermore, the current in the photon-induced side-peaks
is strongly enhanced by increasing electron-photon coupling strength in the $x$-polarized of the photon field,
while a very slightly enhancement in the photon-induced side-peak current was observed in the $y$-polarization photon.
This discrepancy between the polarizations is explained by the anisotropy of the confinement of the
central system. 

Change in the photon-electron coupling strength alters the inter-dot tunneling sensitively,
altering the conduction through the system. To describe this effect properly it is important
to include many higher energy states in the system. Along the similar line of thought our
calculations show that in the present system photon processes of more than one photon are
important.  

The fact that we include both the para- and diamagnetic terms in the electron-photon interaction
leads to complex photon-electron processes that all contribute to the PAT resonance peaks observed.
This has to be viewed in light of the common practice to use only the paramagnetic part in two-level
systems in order to calculate PAT phenomena. In many systems the geometry matters, and the both
parts of the interaction can be important for strong enough coupling.

\ \\
\begin{acknowledgments}
This work was financially supported by the Icelandic Research and Instruments Funds,
the Research Fund of the University of Iceland, the Nordic High Performance Computing facility in Iceland,
and the National Science Council in Taiwan through Contract No.\ NSC100-2112-M-239-001-MY3.
\end{acknowledgments}

%

\bibliographystyle{apsrev4-1}

\begin{thebibliography}{34}%
\makeatletter
\providecommand \@ifxundefined [1]{%
 \@ifx{#1\undefined}
}%
\providecommand \@ifnum [1]{%
 \ifnum #1\expandafter \@firstoftwo
 \else \expandafter \@secondoftwo
 \fi
}%
\providecommand \@ifx [1]{%
 \ifx #1\expandafter \@firstoftwo
 \else \expandafter \@secondoftwo
 \fi
}%
\providecommand \natexlab [1]{#1}%
\providecommand \enquote  [1]{``#1''}%
\providecommand \bibnamefont  [1]{#1}%
\providecommand \bibfnamefont [1]{#1}%
\providecommand \citenamefont [1]{#1}%
\providecommand \href@noop [0]{\@secondoftwo}%
\providecommand \href [0]{\begingroup \@sanitize@url \@href}%
\providecommand \@href[1]{\@@startlink{#1}\@@href}%
\providecommand \@@href[1]{\endgroup#1\@@endlink}%
\providecommand \@sanitize@url [0]{\catcode `\\12\catcode `\$12\catcode
  `\&12\catcode `\#12\catcode `\^12\catcode `\_12\catcode `\%12\relax}%
\providecommand \@@startlink[1]{}%
\providecommand \@@endlink[0]{}%
\providecommand \url  [0]{\begingroup\@sanitize@url \@url }%
\providecommand \@url [1]{\endgroup\@href {#1}{\urlprefix }}%
\providecommand \urlprefix  [0]{URL }%
\providecommand \Eprint [0]{\href }%
\providecommand \doibase [0]{http://dx.doi.org/}%
\providecommand \selectlanguage [0]{\@gobble}%
\providecommand \bibinfo  [0]{\@secondoftwo}%
\providecommand \bibfield  [0]{\@secondoftwo}%
\providecommand \translation [1]{[#1]}%
\providecommand \BibitemOpen [0]{}%
\providecommand \bibitemStop [0]{}%
\providecommand \bibitemNoStop [0]{.\EOS\space}%
\providecommand \EOS [0]{\spacefactor3000\relax}%
\providecommand \BibitemShut  [1]{\csname bibitem#1\endcsname}%
\let\auto@bib@innerbib\@empty
\bibitem [{\citenamefont {Wailfred}\ \emph {et~al.}(2001)\citenamefont
  {Wailfred}, \citenamefont {Wiel}, \citenamefont {Fujisawa}, \citenamefont
  {Tarucham},\ and\ \citenamefont {Kouwenhoven}}]{Kouwenhoven40.2001}%
  \BibitemOpen
  \bibfield  {author} {\bibinfo {author} {\bibfnamefont {G.}~\bibnamefont
  {Wailfred}}, \bibinfo {author} {\bibfnamefont {v.~d.}\ \bibnamefont {Wiel}},
  \bibinfo {author} {\bibfnamefont {T.}~\bibnamefont {Fujisawa}}, \bibinfo
  {author} {\bibfnamefont {S.}~\bibnamefont {Tarucham}}, \ and\ \bibinfo
  {author} {\bibfnamefont {L.~P.}\ \bibnamefont {Kouwenhoven}},\ }\href@noop {}
  {\bibfield  {journal} {\bibinfo  {journal} {Jpn. J. Appl. Phys.}\ }\textbf
  {\bibinfo {volume} {40}},\ \bibinfo {pages} {2100} (\bibinfo {year}
  {2001})}\BibitemShut {NoStop}%
\bibitem [{\citenamefont {van~der Wiel}\ \emph {et~al.}(2003)\citenamefont
  {van~der Wiel}, \citenamefont {{De Franceschi}}, \citenamefont {Elzerman},
  \citenamefont {Fujisawa}, \citenamefont {Tarucham},\ and\ \citenamefont
  {Kouwenhoven}}]{Kouwenhoven.75.2003}%
  \BibitemOpen
  \bibfield  {author} {\bibinfo {author} {\bibfnamefont {W.~G.}\ \bibnamefont
  {van~der Wiel}}, \bibinfo {author} {\bibfnamefont {S.}~\bibnamefont {{De
  Franceschi}}}, \bibinfo {author} {\bibfnamefont {J.~M.}\ \bibnamefont
  {Elzerman}}, \bibinfo {author} {\bibfnamefont {T.}~\bibnamefont {Fujisawa}},
  \bibinfo {author} {\bibfnamefont {S.}~\bibnamefont {Tarucham}}, \ and\
  \bibinfo {author} {\bibfnamefont {L.~P.}\ \bibnamefont {Kouwenhoven}},\
  }\href@noop {} {\bibfield  {journal} {\bibinfo  {journal} {Reviews of modern
  physics}\ }\textbf {\bibinfo {volume} {75}} (\bibinfo {year}
  {2003})}\BibitemShut {NoStop}%
\bibitem [{\citenamefont {Abdullah}\ \emph {et~al.}(2013)\citenamefont
  {Abdullah}, \citenamefont {Tang}, \citenamefont {Manolescu},\ and\
  \citenamefont {Gudmundsson}}]{Nzar.25.465302}%
  \BibitemOpen
  \bibfield  {author} {\bibinfo {author} {\bibfnamefont {N.~R.}\ \bibnamefont
  {Abdullah}}, \bibinfo {author} {\bibfnamefont {C.~S.}\ \bibnamefont {Tang}},
  \bibinfo {author} {\bibfnamefont {A.}~\bibnamefont {Manolescu}}, \ and\
  \bibinfo {author} {\bibfnamefont {V.}~\bibnamefont {Gudmundsson}},\
  }\href@noop {} {\bibfield  {journal} {\bibinfo  {journal} {Journal of
  Physics:Condensed Matter}\ }\textbf {\bibinfo {volume} {25}},\ \bibinfo
  {pages} {465302} (\bibinfo {year} {2013})}\BibitemShut {NoStop}%
\bibitem [{\citenamefont {{Bagheri Tagani}}\ and\ \citenamefont {{Rahimpour
  Soleimani}}(2013)}]{Tagani413:86_91}%
  \BibitemOpen
  \bibfield  {author} {\bibinfo {author} {\bibfnamefont {M.}~\bibnamefont
  {{Bagheri Tagani}}}\ and\ \bibinfo {author} {\bibfnamefont {H.}~\bibnamefont
  {{Rahimpour Soleimani}}},\ }\href@noop {} {\bibfield  {journal} {\bibinfo
  {journal} {Physica B: Condensed Matter}\ }\textbf {\bibinfo {volume} {413}},\
  \bibinfo {pages} {86:91} (\bibinfo {year} {2013})}\BibitemShut {NoStop}%
\bibitem [{\citenamefont {Xu}\ and\ \citenamefont
  {Vavilov}(2013)}]{Xu.87.035429}%
  \BibitemOpen
  \bibfield  {author} {\bibinfo {author} {\bibfnamefont {C.}~\bibnamefont
  {Xu}}\ and\ \bibinfo {author} {\bibfnamefont {M.~G.}\ \bibnamefont
  {Vavilov}},\ }\href {\doibase 10.1103/PhysRevB.87.035429} {\bibfield
  {journal} {\bibinfo  {journal} {Phys. Rev. B}\ }\textbf {\bibinfo {volume}
  {87}},\ \bibinfo {pages} {035429} (\bibinfo {year} {2013})}\BibitemShut
  {NoStop}%
\bibitem [{\citenamefont {Souza}\ \emph {et~al.}(2011)\citenamefont {Souza},
  \citenamefont {Carrara},\ and\ \citenamefont {Vernek}}]{Souza.84.115322}%
  \BibitemOpen
  \bibfield  {author} {\bibinfo {author} {\bibfnamefont {F.~M.}\ \bibnamefont
  {Souza}}, \bibinfo {author} {\bibfnamefont {T.~L.}\ \bibnamefont {Carrara}},
  \ and\ \bibinfo {author} {\bibfnamefont {E.}~\bibnamefont {Vernek}},\ }\href
  {\doibase 10.1103/PhysRevB.84.115322} {\bibfield  {journal} {\bibinfo
  {journal} {Phys. Rev. B}\ }\textbf {\bibinfo {volume} {84}},\ \bibinfo
  {pages} {115322} (\bibinfo {year} {2011})}\BibitemShut {NoStop}%
\bibitem [{\citenamefont {W\"{a}tzel}\ \emph {et~al.}(2011)\citenamefont
  {W\"{a}tzel}, \citenamefont {Moskalenko},\ and\ \citenamefont
  {Berakdar}}]{Watzel99.192101}%
  \BibitemOpen
  \bibfield  {author} {\bibinfo {author} {\bibfnamefont {J.}~\bibnamefont
  {W\"{a}tzel}}, \bibinfo {author} {\bibfnamefont {A.~S.}\ \bibnamefont
  {Moskalenko}}, \ and\ \bibinfo {author} {\bibfnamefont {J.}~\bibnamefont
  {Berakdar}},\ }\href@noop {} {\bibfield  {journal} {\bibinfo  {journal}
  {Appl. Phys. Lett.}\ }\textbf {\bibinfo {volume} {99}},\ \bibinfo {pages}
  {192101} (\bibinfo {year} {2011})}\BibitemShut {NoStop}%
\bibitem [{\citenamefont {Mu\~{n}oz Matutano}\ \emph
  {et~al.}(2011)\citenamefont {Mu\~{n}oz Matutano}, \citenamefont {Royo},
  \citenamefont {Climente}, \citenamefont {Canet-Ferrer}, \citenamefont
  {Fuster}, \citenamefont {Alonso-Gonz\'{a}lez}, \citenamefont
  {Fern\'{a}ndez-Mart\'{\i}nez}, \citenamefont {Mart\'{\i}nez-Pastor},
  \citenamefont {Gonz\'{a}lez}, \citenamefont {Gonz\'{a}lez}, \citenamefont
  {Briones},\ and\ \citenamefont {Al\'{e}n}}]{Matutano84:041308}%
  \BibitemOpen
  \bibfield  {author} {\bibinfo {author} {\bibfnamefont {G.}~\bibnamefont
  {Mu\~{n}oz Matutano}}, \bibinfo {author} {\bibfnamefont {M.}~\bibnamefont
  {Royo}}, \bibinfo {author} {\bibfnamefont {J.~I.}\ \bibnamefont {Climente}},
  \bibinfo {author} {\bibfnamefont {J.}~\bibnamefont {Canet-Ferrer}}, \bibinfo
  {author} {\bibfnamefont {D.}~\bibnamefont {Fuster}}, \bibinfo {author}
  {\bibfnamefont {P.}~\bibnamefont {Alonso-Gonz\'{a}lez}}, \bibinfo {author}
  {\bibfnamefont {I.}~\bibnamefont {Fern\'{a}ndez-Mart\'{\i}nez}}, \bibinfo
  {author} {\bibfnamefont {J.}~\bibnamefont {Mart\'{\i}nez-Pastor}}, \bibinfo
  {author} {\bibfnamefont {Y.}~\bibnamefont {Gonz\'{a}lez}}, \bibinfo {author}
  {\bibfnamefont {L.}~\bibnamefont {Gonz\'{a}lez}}, \bibinfo {author}
  {\bibfnamefont {F.}~\bibnamefont {Briones}}, \ and\ \bibinfo {author}
  {\bibfnamefont {B.}~\bibnamefont {Al\'{e}n}},\ }\href {\doibase
  10.1103/PhysRevB.84.041308} {\bibfield  {journal} {\bibinfo  {journal} {Phys.
  Rev. B}\ }\textbf {\bibinfo {volume} {84}},\ \bibinfo {pages} {041308}
  (\bibinfo {year} {2011})}\BibitemShut {NoStop}%
\bibitem [{\citenamefont {Kaczmarkiewicz}\ \emph {et~al.}(2013)\citenamefont
  {Kaczmarkiewicz}, \citenamefont {Machnikowski},\ and\ \citenamefont
  {Kuhn}}]{Kaczmarkiewicz114:183108}%
  \BibitemOpen
  \bibfield  {author} {\bibinfo {author} {\bibfnamefont {P.}~\bibnamefont
  {Kaczmarkiewicz}}, \bibinfo {author} {\bibfnamefont {P.}~\bibnamefont
  {Machnikowski}}, \ and\ \bibinfo {author} {\bibfnamefont {T.}~\bibnamefont
  {Kuhn}},\ }\href@noop {} {\bibfield  {journal} {\bibinfo  {journal} {Journal
  of Applied Physics}\ }\textbf {\bibinfo {volume} {114}},\ \bibinfo {pages}
  {183108} (\bibinfo {year} {2013})}\BibitemShut {NoStop}%
\bibitem [{\citenamefont {Stafford}\ and\ \citenamefont
  {Wingreen}(1996)}]{Phys.Rev.Lett.76.1996}%
  \BibitemOpen
  \bibfield  {author} {\bibinfo {author} {\bibfnamefont {C.~A.}\ \bibnamefont
  {Stafford}}\ and\ \bibinfo {author} {\bibfnamefont {N.~S.}\ \bibnamefont
  {Wingreen}},\ }\href@noop {} {\bibfield  {journal} {\bibinfo  {journal}
  {Phys. Rev. Lett.}\ }\textbf {\bibinfo {volume} {76}},\ \bibinfo {pages}
  {1916} (\bibinfo {year} {1996})}\BibitemShut {NoStop}%
\bibitem [{\citenamefont {Stoof}\ and\ \citenamefont
  {Nazarov}(1996)}]{PhysRevB.53.1050}%
  \BibitemOpen
  \bibfield  {author} {\bibinfo {author} {\bibfnamefont {T.~H.}\ \bibnamefont
  {Stoof}}\ and\ \bibinfo {author} {\bibfnamefont {Y.~V.}\ \bibnamefont
  {Nazarov}},\ }\href@noop {} {\bibfield  {journal} {\bibinfo  {journal} {Phys.
  Rev. B}\ }\textbf {\bibinfo {volume} {53}},\ \bibinfo {pages} {1050}
  (\bibinfo {year} {1996})}\BibitemShut {NoStop}%
\bibitem [{\citenamefont {Shang}\ \emph {et~al.}(2013)\citenamefont {Shang},
  \citenamefont {Li}, \citenamefont {Gao}, \citenamefont {Xiao}, \citenamefont
  {Tu}, \citenamefont {Jiang}, \citenamefont {Guo},\ and\ \citenamefont
  {Guo}}]{Shang.103.2013}%
  \BibitemOpen
  \bibfield  {author} {\bibinfo {author} {\bibfnamefont {R.}~\bibnamefont
  {Shang}}, \bibinfo {author} {\bibfnamefont {H.-O.}\ \bibnamefont {Li}},
  \bibinfo {author} {\bibfnamefont {G.}~\bibnamefont {Gao}}, \bibinfo {author}
  {\bibfnamefont {M.}~\bibnamefont {Xiao}}, \bibinfo {author} {\bibfnamefont
  {T.}~\bibnamefont {Tu}}, \bibinfo {author} {\bibfnamefont {H.}~\bibnamefont
  {Jiang}}, \bibinfo {author} {\bibfnamefont {G.-C.}\ \bibnamefont {Guo}}, \
  and\ \bibinfo {author} {\bibfnamefont {G.-P.}\ \bibnamefont {Guo}},\
  }\href@noop {} {\bibfield  {journal} {\bibinfo  {journal} {Appl. Phys.
  Lett.}\ }\textbf {\bibinfo {volume} {103}},\ \bibinfo {pages} {162109}
  (\bibinfo {year} {2013})}\BibitemShut {NoStop}%
\bibitem [{\citenamefont {Shibata}\ \emph {et~al.}(2012)\citenamefont
  {Shibata}, \citenamefont {Umeno}, \citenamefont {Cha},\ and\ \citenamefont
  {Hirakawa}}]{Shibata109.077401}%
  \BibitemOpen
  \bibfield  {author} {\bibinfo {author} {\bibfnamefont {K.}~\bibnamefont
  {Shibata}}, \bibinfo {author} {\bibfnamefont {A.}~\bibnamefont {Umeno}},
  \bibinfo {author} {\bibfnamefont {K.~M.}\ \bibnamefont {Cha}}, \ and\
  \bibinfo {author} {\bibfnamefont {K.}~\bibnamefont {Hirakawa}},\ }\href
  {\doibase 10.1103/PhysRevLett.109.077401} {\bibfield  {journal} {\bibinfo
  {journal} {Phys. Rev. Lett.}\ }\textbf {\bibinfo {volume} {109}},\ \bibinfo
  {pages} {077401} (\bibinfo {year} {2012})}\BibitemShut {NoStop}%
\bibitem [{\citenamefont {Imamoglu}\ and\ \citenamefont
  {Yamamoto}(1994)}]{Imamog72.210}%
  \BibitemOpen
  \bibfield  {author} {\bibinfo {author} {\bibfnamefont {A.}~\bibnamefont
  {Imamoglu}}\ and\ \bibinfo {author} {\bibfnamefont {Y.}~\bibnamefont
  {Yamamoto}},\ }\href {\doibase 10.1103/PhysRevLett.72.210} {\bibfield
  {journal} {\bibinfo  {journal} {Phys. Rev. Lett.}\ }\textbf {\bibinfo
  {volume} {72}},\ \bibinfo {pages} {210} (\bibinfo {year} {1994})}\BibitemShut
  {NoStop}%
\bibitem [{\citenamefont {Loss}\ and\ \citenamefont
  {DiVincenzo}(1998)}]{Loss.57.1998}%
  \BibitemOpen
  \bibfield  {author} {\bibinfo {author} {\bibfnamefont {D.}~\bibnamefont
  {Loss}}\ and\ \bibinfo {author} {\bibfnamefont {D.~P.}\ \bibnamefont
  {DiVincenzo}},\ }\href@noop {} {\bibfield  {journal} {\bibinfo  {journal}
  {Physical Review A}\ }\textbf {\bibinfo {volume} {57}},\ \bibinfo {pages}
  {120} (\bibinfo {year} {1998})}\BibitemShut {NoStop}%
\bibitem [{\citenamefont {DiVincenzo}(2005)}]{DiVincenzo.309.2005}%
  \BibitemOpen
  \bibfield  {author} {\bibinfo {author} {\bibfnamefont {D.~P.}\ \bibnamefont
  {DiVincenzo}},\ }\href@noop {} {\bibfield  {journal} {\bibinfo  {journal}
  {Science}\ }\textbf {\bibinfo {volume} {309}} (\bibinfo {year}
  {2005})}\BibitemShut {NoStop}%
\bibitem [{\citenamefont {Chuang}\ and\ \citenamefont
  {Nielsen}(2010)}]{Nielsen.2010}%
  \BibitemOpen
  \bibfield  {author} {\bibinfo {author} {\bibfnamefont {L.~I.}\ \bibnamefont
  {Chuang}}\ and\ \bibinfo {author} {\bibfnamefont {M.~A.}\ \bibnamefont
  {Nielsen}},\ }\href@noop {} {\bibfield  {journal} {\bibinfo  {journal}
  {\textit{Quantum Computation and Quantum Information}}\ } (\bibinfo {year}
  {Cambridge University Press, 2010})}\BibitemShut {NoStop}%
\bibitem [{\citenamefont {Vaz}\ and\ \citenamefont
  {Kyriakidis}(2010)}]{Vaz81.085315}%
  \BibitemOpen
  \bibfield  {author} {\bibinfo {author} {\bibfnamefont {E.}~\bibnamefont
  {Vaz}}\ and\ \bibinfo {author} {\bibfnamefont {J.}~\bibnamefont
  {Kyriakidis}},\ }\href {\doibase 10.1103/PhysRevB.81.085315} {\bibfield
  {journal} {\bibinfo  {journal} {Phys. Rev. B}\ }\textbf {\bibinfo {volume}
  {81}},\ \bibinfo {pages} {085315} (\bibinfo {year} {2010})}\BibitemShut
  {NoStop}%
\bibitem [{\citenamefont {Gurvitz}\ and\ \citenamefont
  {Prager}(1996)}]{Gurvitz53.15932}%
  \BibitemOpen
  \bibfield  {author} {\bibinfo {author} {\bibfnamefont {S.~A.}\ \bibnamefont
  {Gurvitz}}\ and\ \bibinfo {author} {\bibfnamefont {Y.~S.}\ \bibnamefont
  {Prager}},\ }\href {\doibase 10.1103/PhysRevB.53.15932} {\bibfield  {journal}
  {\bibinfo  {journal} {Phys. Rev. B}\ }\textbf {\bibinfo {volume} {53}},\
  \bibinfo {pages} {15932} (\bibinfo {year} {1996})}\BibitemShut {NoStop}%
\bibitem [{\citenamefont {Harbola}\ \emph {et~al.}(2006)\citenamefont
  {Harbola}, \citenamefont {Esposito},\ and\ \citenamefont
  {Mukamel}}]{Harbola74.235309}%
  \BibitemOpen
  \bibfield  {author} {\bibinfo {author} {\bibfnamefont {U.}~\bibnamefont
  {Harbola}}, \bibinfo {author} {\bibfnamefont {M.}~\bibnamefont {Esposito}}, \
  and\ \bibinfo {author} {\bibfnamefont {S.}~\bibnamefont {Mukamel}},\ }\href
  {\doibase 10.1103/PhysRevB.74.235309} {\bibfield  {journal} {\bibinfo
  {journal} {Phys. Rev. B}\ }\textbf {\bibinfo {volume} {74}},\ \bibinfo
  {pages} {235309} (\bibinfo {year} {2006})}\BibitemShut {NoStop}%
\bibitem [{\citenamefont {{Van Kampen}}(2001)}]{Kampen2.2001}%
  \BibitemOpen
  \bibfield  {author} {\bibinfo {author} {\bibfnamefont {N.~G.}\ \bibnamefont
  {{Van Kampen}}},\ }\href@noop {} {\bibfield  {journal} {\bibinfo  {journal}
  {\textit{Stochastic Processes in Physics and Chemistry}}\ }\textbf {\bibinfo
  {volume} {2nd Ed}} (\bibinfo {year} {North-Holland, Amsterdam,
  2001})}\BibitemShut {NoStop}%
\bibitem [{\citenamefont {Bednorz}\ and\ \citenamefont
  {Belzig}(2008)}]{Bednorz101.206803}%
  \BibitemOpen
  \bibfield  {author} {\bibinfo {author} {\bibfnamefont {A.}~\bibnamefont
  {Bednorz}}\ and\ \bibinfo {author} {\bibfnamefont {W.}~\bibnamefont
  {Belzig}},\ }\href {\doibase 10.1103/PhysRevLett.101.206803} {\bibfield
  {journal} {\bibinfo  {journal} {Phys. Rev. Lett.}\ }\textbf {\bibinfo
  {volume} {101}},\ \bibinfo {pages} {206803} (\bibinfo {year}
  {2008})}\BibitemShut {NoStop}%
\bibitem [{\citenamefont {Braggio}\ \emph {et~al.}(2006)\citenamefont
  {Braggio}, \citenamefont {K\"{o}nig},\ and\ \citenamefont
  {Fazio}}]{Braggio96.026805}%
  \BibitemOpen
  \bibfield  {author} {\bibinfo {author} {\bibfnamefont {A.}~\bibnamefont
  {Braggio}}, \bibinfo {author} {\bibfnamefont {J.}~\bibnamefont {K\"{o}nig}},
  \ and\ \bibinfo {author} {\bibfnamefont {R.}~\bibnamefont {Fazio}},\ }\href
  {\doibase 10.1103/PhysRevLett.96.026805} {\bibfield  {journal} {\bibinfo
  {journal} {Phys. Rev. Lett.}\ }\textbf {\bibinfo {volume} {96}},\ \bibinfo
  {pages} {026805} (\bibinfo {year} {2006})}\BibitemShut {NoStop}%
\bibitem [{\citenamefont {Emary}\ \emph {et~al.}(2007)\citenamefont {Emary},
  \citenamefont {Marcos}, \citenamefont {Aguado},\ and\ \citenamefont
  {Brandes}}]{Emary76.161404}%
  \BibitemOpen
  \bibfield  {author} {\bibinfo {author} {\bibfnamefont {C.}~\bibnamefont
  {Emary}}, \bibinfo {author} {\bibfnamefont {D.}~\bibnamefont {Marcos}},
  \bibinfo {author} {\bibfnamefont {R.}~\bibnamefont {Aguado}}, \ and\ \bibinfo
  {author} {\bibfnamefont {T.}~\bibnamefont {Brandes}},\ }\href {\doibase
  10.1103/PhysRevB.76.161404} {\bibfield  {journal} {\bibinfo  {journal} {Phys.
  Rev. B}\ }\textbf {\bibinfo {volume} {76}},\ \bibinfo {pages} {161404}
  (\bibinfo {year} {2007})}\BibitemShut {NoStop}%
\bibitem [{\citenamefont {Gudmundsson}\ \emph {et~al.}(2009)\citenamefont
  {Gudmundsson}, \citenamefont {Gainar}, \citenamefont {Tang}, \citenamefont
  {Moldoveanu},\ and\ \citenamefont {Manolecu}}]{Vidar11.113007}%
  \BibitemOpen
  \bibfield  {author} {\bibinfo {author} {\bibfnamefont {V.}~\bibnamefont
  {Gudmundsson}}, \bibinfo {author} {\bibfnamefont {C.}~\bibnamefont {Gainar}},
  \bibinfo {author} {\bibfnamefont {C.-S.}\ \bibnamefont {Tang}}, \bibinfo
  {author} {\bibfnamefont {V.}~\bibnamefont {Moldoveanu}}, \ and\ \bibinfo
  {author} {\bibfnamefont {A.}~\bibnamefont {Manolecu}},\ }\href@noop {}
  {\bibfield  {journal} {\bibinfo  {journal} {New J. Phys.}\ }\textbf {\bibinfo
  {volume} {11}},\ \bibinfo {pages} {113007} (\bibinfo {year}
  {2009})}\BibitemShut {NoStop}%
\bibitem [{\citenamefont {Abdullah}\ \emph {et~al.}(2010)\citenamefont
  {Abdullah}, \citenamefont {Tang},\ and\ \citenamefont
  {Gudmundsson}}]{Abdullah52.195325}%
  \BibitemOpen
  \bibfield  {author} {\bibinfo {author} {\bibfnamefont {N.~R.}\ \bibnamefont
  {Abdullah}}, \bibinfo {author} {\bibfnamefont {C.-S.}\ \bibnamefont {Tang}},
  \ and\ \bibinfo {author} {\bibfnamefont {V.}~\bibnamefont {Gudmundsson}},\
  }\href {\doibase 10.1103/PhysRevB.82.195325} {\bibfield  {journal} {\bibinfo
  {journal} {Phys. Rev. B}\ }\textbf {\bibinfo {volume} {82}},\ \bibinfo
  {pages} {195325} (\bibinfo {year} {2010})}\BibitemShut {NoStop}%
\bibitem [{\citenamefont {Yannouleas}\ and\ \citenamefont
  {Landman}(2007)}]{Yannouleas70.2067}%
  \BibitemOpen
  \bibfield  {author} {\bibinfo {author} {\bibfnamefont {C.}~\bibnamefont
  {Yannouleas}}\ and\ \bibinfo {author} {\bibfnamefont {U.}~\bibnamefont
  {Landman}},\ }\href@noop {} {\bibfield  {journal} {\bibinfo  {journal} {Rep.
  Prog. Phys}\ }\textbf {\bibinfo {volume} {70}},\ \bibinfo {pages} {2067}
  (\bibinfo {year} {2007})}\BibitemShut {NoStop}%
\bibitem [{\citenamefont {Gudmundsson}\ \emph {et~al.}(2012)\citenamefont
  {Gudmundsson}, \citenamefont {Jonasson}, \citenamefont {Tang}, \citenamefont
  {Goan},\ and\ \citenamefont {Manolescu}}]{Vidar85.075306}%
  \BibitemOpen
  \bibfield  {author} {\bibinfo {author} {\bibfnamefont {V.}~\bibnamefont
  {Gudmundsson}}, \bibinfo {author} {\bibfnamefont {O.}~\bibnamefont
  {Jonasson}}, \bibinfo {author} {\bibfnamefont {C.-S.}\ \bibnamefont {Tang}},
  \bibinfo {author} {\bibfnamefont {H.-S.}\ \bibnamefont {Goan}}, \ and\
  \bibinfo {author} {\bibfnamefont {A.}~\bibnamefont {Manolescu}},\ }\href
  {\doibase 10.1103/PhysRevB.85.075306} {\bibfield  {journal} {\bibinfo
  {journal} {Phys. Rev. B}\ }\textbf {\bibinfo {volume} {85}},\ \bibinfo
  {pages} {075306} (\bibinfo {year} {2012})}\BibitemShut {NoStop}%
\bibitem [{\citenamefont {Gudmundsson}\ \emph {et~al.}(2013)\citenamefont
  {Gudmundsson}, \citenamefont {Jonasson}, \citenamefont {Arnold},
  \citenamefont {Tang}, \citenamefont {Goan},\ and\ \citenamefont
  {Manolescu}}]{Vidar61.305}%
  \BibitemOpen
  \bibfield  {author} {\bibinfo {author} {\bibfnamefont {V.}~\bibnamefont
  {Gudmundsson}}, \bibinfo {author} {\bibfnamefont {O.}~\bibnamefont
  {Jonasson}}, \bibinfo {author} {\bibfnamefont {T.}~\bibnamefont {Arnold}},
  \bibinfo {author} {\bibfnamefont {C.-S.}\ \bibnamefont {Tang}}, \bibinfo
  {author} {\bibfnamefont {H.-S.}\ \bibnamefont {Goan}}, \ and\ \bibinfo
  {author} {\bibfnamefont {A.}~\bibnamefont {Manolescu}},\ }\href@noop {}
  {\bibfield  {journal} {\bibinfo  {journal} {Fortschr. Phys.}\ }\textbf
  {\bibinfo {volume} {61}},\ \bibinfo {pages} {305} (\bibinfo {year}
  {2013})}\BibitemShut {NoStop}%
\bibitem [{\citenamefont {Jin}\ \emph {et~al.}(2008)\citenamefont {Jin},
  \citenamefont {Zheng},\ and\ \citenamefont {Yan}}]{Jinshuang128.1234703}%
  \BibitemOpen
  \bibfield  {author} {\bibinfo {author} {\bibfnamefont {J.~S.}\ \bibnamefont
  {Jin}}, \bibinfo {author} {\bibfnamefont {X.}~\bibnamefont {Zheng}}, \ and\
  \bibinfo {author} {\bibfnamefont {Y.}~\bibnamefont {Yan}},\ }\href@noop {}
  {\bibfield  {journal} {\bibinfo  {journal} {J. Chem. Phys. Lett.}\ }\textbf
  {\bibinfo {volume} {128}},\ \bibinfo {pages} {234703} (\bibinfo {year}
  {2008})}\BibitemShut {NoStop}%
\bibitem [{\citenamefont {Haake}(1971)}]{Haake3.1723}%
  \BibitemOpen
  \bibfield  {author} {\bibinfo {author} {\bibfnamefont {F.}~\bibnamefont
  {Haake}},\ }\href {\doibase 10.1103/PhysRevA.3.1723} {\bibfield  {journal}
  {\bibinfo  {journal} {Phys. Rev. A}\ }\textbf {\bibinfo {volume} {3}},\
  \bibinfo {pages} {1723} (\bibinfo {year} {1971})}\BibitemShut {NoStop}%
\bibitem [{\citenamefont {Breuer}\ and\ \citenamefont
  {Petruccione}(2002)}]{Breuer2002}%
  \BibitemOpen
  \bibfield  {author} {\bibinfo {author} {\bibfnamefont {H.-P.}\ \bibnamefont
  {Breuer}}\ and\ \bibinfo {author} {\bibfnamefont {F.}~\bibnamefont
  {Petruccione}},\ }\href@noop {} {\bibfield  {journal} {\bibinfo  {journal}
  {\textit{The Theory of Open Quantum Systems}}\ } (\bibinfo {year} {Oxford
  University Press, Oxford, 2002})}\BibitemShut {NoStop}%
\bibitem [{\citenamefont {Kouwenhoven}\ \emph {et~al.}(1994)\citenamefont
  {Kouwenhoven}, \citenamefont {Jauhar}, \citenamefont {Orenstein},
  \citenamefont {McEuen}, \citenamefont {Nagamune}, \citenamefont {Motohisa},\
  and\ \citenamefont {Sakaki}}]{Kouwenhoven73.3443}%
  \BibitemOpen
  \bibfield  {author} {\bibinfo {author} {\bibfnamefont {L.~P.}\ \bibnamefont
  {Kouwenhoven}}, \bibinfo {author} {\bibfnamefont {S.}~\bibnamefont {Jauhar}},
  \bibinfo {author} {\bibfnamefont {J.}~\bibnamefont {Orenstein}}, \bibinfo
  {author} {\bibfnamefont {P.~L.}\ \bibnamefont {McEuen}}, \bibinfo {author}
  {\bibfnamefont {Y.}~\bibnamefont {Nagamune}}, \bibinfo {author}
  {\bibfnamefont {J.}~\bibnamefont {Motohisa}}, \ and\ \bibinfo {author}
  {\bibfnamefont {H.}~\bibnamefont {Sakaki}},\ }\href {\doibase
  10.1103/PhysRevLett.73.3443} {\bibfield  {journal} {\bibinfo  {journal}
  {Phys. Rev. Lett.}\ }\textbf {\bibinfo {volume} {73}},\ \bibinfo {pages}
  {3443} (\bibinfo {year} {1994})}\BibitemShut {NoStop}%
\bibitem [{\citenamefont {Platero}\ and\ \citenamefont
  {Aguado}(2004)}]{Platero.395.2004}%
  \BibitemOpen
  \bibfield  {author} {\bibinfo {author} {\bibfnamefont {G.}~\bibnamefont
  {Platero}}\ and\ \bibinfo {author} {\bibfnamefont {R.}~\bibnamefont
  {Aguado}},\ }\href@noop {} {\bibfield  {journal} {\bibinfo  {journal}
  {Physics Report}\ }\textbf {\bibinfo {volume} {395}},\ \bibinfo {pages} {1}
  (\bibinfo {year} {2004})}\BibitemShut {NoStop}%
\end{thebibliography}

%
%
\end{document}